\documentclass{article}
\newcommand\vv[1]{{\overline{#1}}}
\newcommand{\Rule}[3]{\frac{\displaystyle #1}{\displaystyle #2} \ \ #3}
\newcommand{\RRule}[3]{\[\Rule{#1}{#2}{#3}\]}
\newtheorem{proposition}{Proposition}[section]
\newtheorem{theorem}{Theorem}
\newtheorem{corollary}{Corollary}

\newtheorem{Define}[proposition]{Definition}

\newlength{\proofbox}
\newenvironment{proof}{\begin{description}\item[Proof:]}{
\hspace*{\fill}\rule{\proofbox}{\proofbox}
\end{description}}
\advance\hoffset by -.7in
\title{{\bf An Abstract Programming System}}
\author{
David A. Plaisted\thanks{This research was partially supported by the
National Science Foundation under grant CCR-9972118.}\\
Department of Computer Science\\
UNC Chapel Hill\\
Chapel Hill, NC 27599-3175\\
Phone: (919) 967-9238\\
Fax: (919) 962-1799 \\
{\small Email: plaisted@cs.unc.edu}}
\begin{document}
\bibliographystyle{alpha}
\maketitle
\begin{abstract}
The system $PL$ permits the translation of abstract proofs of program
correctness into programs in a variety of programming languages.  A
programming language satisfying certain axioms may be the target of
such a translation.  The system $PL$ also permits the construction and
proof of correctness of programs in an abstract programming language,
and permits the translation of these programs into correct programs in
a variety of languages.  The abstract programming language has an
imperative style of programming with assignment statements and
side-effects, to allow the efficient generation of code.  The abstract
programs may be written by humans and then translated, avoiding the
need to write the same program repeatedly in different languages or
even the same language.  This system uses classical logic, is
conceptually simple, and permits reasoning about nonterminating
programs using Scott-Strachey style denotational semantics.
\end{abstract}
\section*{Keywords}
Abstract programming, program generation, program verification

\section{Introduction}
The purpose of the system $PL$ is to permit the construction of proofs
that can be viewed as abstract programs and translated into correct
programs in a variety of programming languages.  The emphasis is not
on the automatic construction of the proofs but on the process of
translating them into programs in specific programming languages.  A
programming language must satisfy certain conditions in order to be
the target of such a translation, and typical procedural, functional,
and logic programming languages satisfy these conditions.  This system
uses classical logic and Scott-Strachey domain theory \cite{Stoy:77}.
This system therefore, in theory, permits reasoning about
nonterminating programs and nondeterministic programs.  The system may
also be used to construct abstract programs without proving them
correct, and these abstract programs can also be translated into a
variety of programming languages.

For a given programming language $L$, the axioms $PL(L)$ describe the
properties the language $L$ must satisfy in order for $PL$ proofs to
be translatable into $L$.  A programming language $L$ is {\em
PL-feasible} if it satisfies the axioms of $PL(L)$.  (Actually, the
set of $PL$-feasible languages may differ from one program to another,
because some programs may require different precisions of floating
point numbers or various sizes of character strings or integers that
may not be available in all languages, et cetera.)  $PL(L)$ permits
the construction of proofs that a particular $L$ program $P$ satisfies
a specification.  The system $PL^*$ permits the construction of
abstract proofs that correspond to $PL(L)$ proofs and therefore to
correct programs in any $PL$-feasible language $L$.  For each
$PL$-feasible language $L$, there is an effective function from proofs
in $PL^*$ to correct programs in $L$.  This permits the translation of
$PL^*$ proofs into correct programs in any $PL$-feasible language.  An
abstract programming language $L^*$ corresponds to the system $PL^*$,
and there are likewise effective functions from correct $L^*$ programs
to correct programs in any $PL$-feasible language $L$.  It is possible
to prove the correctness of $L^*$ programs using $PL^*$ or instead to
gain confidence in the reliability of $L^*$ programs by testing or
some other means; thus it is not necessary to prove correctness in
order to translate $L^*$ programs into $PL$-feasible languages.  Note
that $PL$ is not concerned with
the details of the semantics of
$PL$-feasible languages $L$; it only requires that $L$ satisfy the
axioms given for $PL$-feasible languages.

In order to have a realistic representation of algorithms at an
abstract level in $PL$, it is necessary to include imperative
features in $PL$
that permit an accurate representation of operations and data
structures that efficient algorithms use.  For example, $PL$ formalizes the
destructive modification of data, as well as the side effects of
operations on data; one cannot realistically describe quicksort
without the former, and one cannot realistically describe binary tree
manipulation routines without the latter.  The operations in $PL$
are few enough in number to preserve its simplicity, but inclusive enough to
permit the generation of efficient code through the specification of
destructive assignment statements and side effects.

\section{Background and Discussion}

There has been substantial work in program generation and programming
logics.  A number of papers discuss program synthesis based on
constructive logic, type theory, and the Curry-Howard isomorphism
\cite{Constable:85,crsh:93}.  For example, Martin-L\"of's theory of
types \cite{Martin-Lof:82} and the calculus of constructions of
Coquand and Huet \cite{cohu:88,vonhenke95construction} are used for
this purpose.  Bittel \cite{Bittel:92} and Kanovich \cite{Kanovich:91}
describe program synthesis in intuitionistic logic.  NuPRL
\cite{nuprl-book} uses a hybrid system of logic and type theory.  The
proofs-as-programs paradigm of Bates and Constable \cite{baco:85}
interprets constructive proofs as executable programs using the
Curry-Howard isomorphism.  However, such proofs contains both
computational content and correctness arguments; in order to obtain
efficient code, it is useful to separate these, which is not always
simple (Berger and Schwichtenberg \cite{BeSc:96}).  Avellone et al
\cite{avfemi:99} discuss this separation in the context of a system
for reasoning about abstract data types.  They discuss program
synthesis from constructive proofs, but in which abstract data types
have classical semantics.  Jeavons et al \cite{jeavons00fred} present
another system for separating these parts of constructive proofs using
the Curry-Howard isomorphism.  Most such systems synthesize functional
programs, but Mason \cite{Mason:97} studies the synthesis of
imperative programs in a lambda calculus framework.

Another line of work in program generation makes use of {\em schemata}.
The idea of schema based program synthesis \cite{fllaorri:99} is to
consider a schema as a generalized program that can be instantiated to
a number of specific programs.  Such schemata can be derived formally
or constructed manually, and the correctness proofs can be manual or
with automated assistance.  Once obtained, schemata can be
instantiated and combined to produce a variety of correct programs.
Also, systematic methods exist for transforming schemata into different
schemata.  Huet and Lang \cite{hula:78} studied the transformation of one
schema to another, and many other such transformation systems
(\cite{partsch:90}) have
been studied.  This can include tail recursion elimination,
for example.  An advantage of the schema-based approach to program
generation as compared to constructive derivations of programs from
scratch is that the deductive and programming tasks are easier.  The
schema-based approach approach also permits the use of classical logic
in reasoning about schemata.  Of course, schema-based development is
also possible in constructive logics.  Anderson and Basin \cite{anba:00}
mention that schemata are not general enough to capture some
programming knowledge, including the design patters of Gamma et al
\cite{gahejovl:95}.

Schema based program synthesis as in Flener et al \cite{fllaorri:99}
is concerned with synthesizing a variety of logic programs, possibly
in a single language, from a schema.  A schema is typically {\em
open}, which means that some of the predicates (representing
procedures) are not defined.  Thus, the schema can instantiate to
different programs if different definitions of the undefined
procedures are given.  Flener et al \cite{fllaorri:99} separate a
schema into a {\em template}, which is an abstract program, and a {\em
specification framework}, or collection of axioms giving the intended
semantics of the problem domain.  Their work was strongly influenced
by the work of Smith \cite{smith90kids} in this respect.  The
semantics can either be {\em isoinitial}, a restriction of initial
algebra semantics in which negative as well as positive equations are
preserved \cite{fllaor:98}, or can be based on logic program semantics
using {\em completions} of a logic program \cite{laortar:99}.  A
schema that is {\em steadfast} is guaranteed always to instantiate to
correct programs.  The synthesis process combines schemata, often by
instantiating the open predicates of one schema using predicates from
another.  B\"uy\"ukyildiz and Flener \cite{bufl:97} study rules for the
transformation of one logic program schema into another.  Lau,
Ornaghi, and T\"arnlund \cite{laortar:99} discuss the relationship of
schemata to object-oriented programming.  Deville and Lau
\cite{deville94logic} discuss constructive, deductive, and inductive
synthesis of logic programs.

Anderson and Basin \cite{anba:00} show how to view program schemata as
derived rules of inference in higher-order logic.  This approach can
encompass both functional and logic programming
languages.  Like the schema approach, this approach relies on
classical logic and does not make initial algebra assumptions that are
typical of abstract data type theory.  The formalism of Anderson and
Basin \cite{anba:00} makes use of {\em schema variables}, which can be
replaced by arbitrary functions.  Also, Anderson and Basin
\cite{anba:00} emphasize logic programs, but remark that schema based
development also applies readily to functional programs.  Shankar
\cite{shankar:96} and Dold \cite{Dold:95} study program transformation
in higher-order logic in the PVS system.

Manna and Waldinger's deductive tableau system
\cite{mawa:92} also uses classical logic for synthesis of
functional programs.  Ayari and Basin\cite{abba:01} show how to express this
system in Isabelle using higher-order logic and higher-order
resolution.  They give an example of synthesizing sorting programs
(including the quicksort program) in a functional language.

The common language runtime \cite{box:02} of Microsoft is an attempt to
ensure compatibility between different programming languages by
compiling them all into a common intermediate language.  This permits
programs in different languages to communicate with each other.

$PL$ has some features in common with the preceding systems.  As with
schemata, $PL$ is based on classical logic, is not based on the
Curry-Howard isomorphism, and is only concerned with type theory in an
incidental way.  $PL$ also explicitly separates correctness arguments
from the computational content of a program.  Abstract $PL$ programs
are similar to schemata, or templates.  An abstract program in $PL$ is
only partially specified, in the sense that some of the procedures it
uses may not be defined.  This corresponds to the {\em open} programs
of Flener et al \cite{fllaorri:99}.  Steadfast logic programs
correspond to a $PL$ program fragment that satisfies a specification.
Abstract programs can be combined in $PL$, as the schemata of Flener et
al \cite{fllaorri:99}, by instantiating the undefined procedures of
one program to the procedures of another and combining the programs.
Thus an abstract $PL$ program can be viewed as a rule of inference for
constructing programs, as in the system of Anderson and
Basin\cite{anba:00}.  The goal of $PL$ is to avoid the need to write
the same program over and over in different languages by permitting
abstract $PL$ programs to translate to many languages.  Also, $PL$ even
avoids the need to write the same program many times in the same language,
because it permits substitutions for the names of the procedures in
an abstract program.

However, the preceding approaches have a different emphasis than $PL$.
$PL$ does not emphasize the underlying logic; any sufficiently
expressive logic, such as some version of set theory or higher-order
logic, would suffice.  In $PL$, a single logical formula mentions both
an abstract formula and the properties it is assumed to have, rather
than separating this information as is often done with schemata.  In
this, our approach is similar to that of Anderson and Basin
\cite{anba:00}.  Furthermore, $PL$ semantics is based on
Scott-Strachey style \cite{Stoy:77} denotational semantics, and
therefore can potentially reason about nonterminating and even
nondeterministic computation.  $PL$ semantics is defined
axiomatically, instead of by initial or iso-initial models.  Also, the
present paper is not concerned with program generation methodology
{\em per se}, as are a number of other works.  In $PL$, the emphasis
is not on synthesis but on translation of an abstract program into
efficient programs in a variety of languages.  This is related to the
research topic mentioned in Anderson and Basin \cite{anba:00} of
developing a metatheory to transfer schema results from one area to
another.  Some other systems handle synthesis in particular
languages including UNIX, object code, and logic programs.  For
example, Sanella and Tarlecki \cite{sannella89toward} discuss the
formal development of ML programs from algebraic specifications.
Bhansali and Harandi \cite{bhansali93synthesis} discuss the synthesis
of UNIX programs.  Benini \cite{Benini:00} discusses program synthesis
of object code.  The process of translation of $PL$ programs into
other languages is automatic and does not require any planning or
reasoning.  In contrast to other systems, $PL$ does not emphasize the
transformation of one schema to another, except for the translation of
schemata into specific languages and the combination of existing
schemata.  Another difference between $PL$ and other approaches is
that $PL$ variables can only be replaced by procedure names, and not
by arbitrary functions as in Anderson and Basin\cite{anba:00}.

In addition, $PL$ gives substantial attention to imperative features
such as assignment statements and side effects that are important for
efficient code generation.  The treatment of side effects is somewhat
similar to that of Mason \cite{Mason:97}.  The example of quicksort
from Ayari and Basin \cite{abba:01} uses a functional notation, in
which array segments are concatenated, instead of the usual, more
efficient approach of in-place processing of subarrays in the
recursive step.  Bornat \cite{bornat00proving} gives a way to prove
properties of pointer programs using Hoare logic.  Another approach to
side effects is given by Harman et al \cite{harman01sideeffect} and
involves transforming programs to remove side effects.  Though $PL$
emphasizes imperative languages, it also has applications to logic and
functional programs.  By comparision, few if any of the preceding
systems emphasize translation into a variety of languages, nor do most
of them emphasize imperative features of languages.  In addition, the
focus of many of these system is the method of program generation.

The focus of $PL$ differs from that of the common language runtime,
as well.  The latter permits programs in different languages to
communicate.  $PL$ permits an abstract program to translate into
a variety of other languages at the source level, and thereby
avoids the need to write the same program many times in many
languages.

In general, $PL$ is not so much concerned with how programs are
synthesized as with axiomatizing their correctness in an abstract
setting, so as to guarantee the correctness of their translations into
specific languages.  The user would typically write programs in $PL$
and provide proofs of their correctness.

Current program generation methods have several problems: 1.  The demands
on the formal reasoning part of the process are too stringent.  2.
The generated code is not always as efficient as possible.  3.  The
logic is often unfamiliar to the typical user.

The system $PL$ seeks to overcome these problems by permitting the
writing and debugging of abstract programs without any reasoning at
all, if desired.  However, it is possible to verify the abstract
programs.  These programs then translate into efficient code in a
variety of other languages.  This is possible because the abstract
programs permit an imperative style of programming with assignment
statements and side effects.  The use of classical logic helps to
solve the third problem.

\section{Axioms of Program Language Semantics}

\subsection{Introduction to Axioms}
The system $PL(L)$ refers to {\em program fragments} in programming
languages $L$.  A
program fragment $P$ of $L$ is a portion of a program in $L$ that
specifies the definition of some procedures and data in terms of
others.  Data may be integers, arrays, lists, trees, or other data
structures typically referenced by program variables.  The {\em outputs}
of $P$ are the procedures and data that are defined in $P$ in terms of
other procedures and data.  The {\em inputs} of $P$ are the procedures
and data that are referenced in $P$ but not defined there.  Thus if
$\vv{x}$ are the inputs of $P$ and $\vv{y}$ are the outputs of $P$,
$P$ defines a function from the semantics of $\vv{x}$ to the
semantics of $\vv{y}$.  $\vv{x}$ and $\vv{y}$ are {\em variables} of
$P$ in the system $PL(L)$.

There are several operations on program fragments in the system
$PL(L)$.  If $P$ and $Q$ are program fragments, then $P;Q$ represents
the sequential composition of $P$ and $Q$ ($P$ then $Q$).  $PL(L)$
does not have a parallel composition operator; it would be possible
to add additional operators
such as parallel composition and object inheritance to
$PL(L)$.  $\exists x P$ represents $P$ with the variable $x$ declared
``local'' so that it is not visible outside of $\exists x P$.  If
$\Theta$ is a substitution, then $P\Theta$ is $P$ with program
variables (procedures and data) substituted as specified by $\Theta$.
$\mu$ is a least fixpoint operator on programs, corresponding to the
definition of recursive procedures.  $\uparrow^p_{\vv{x}}P$ represents
$P$ with the new procedure $p$ defined; this corresponds to a program
of the form {\bf procedure} $p(\vv{x}); P$ having $P$ as the procedure
body.  $\downarrow^p_{\vv{x}}P$ represents the application of a
procedure to arguments.  This corresponds to a program of the form $P;
\mbox{ {\bf call }} p(\vv{x})$ where $p$ is a procedure defined in
$P$.  The program $P$ may be empty in this case.  $x ? P ? Q$
represents the conditional ``{\bf if} $x$ {\bf then} $P$ {\bf else}
$Q$'' where $P$ and $Q$ are program fragments, possibly empty.
$\mu(u,v,P,\geq)$ represents the ``fixed point'' of $P$ with
procedures $u$ and $v$ identified, where $\geq$ is the ``definedness''
ordering for the denotational semantics of $P$.  Each program $P$ has
a corresponding {\em $PL$ textual syntax} $P^{text}$ so that the
textual syntax for $P;Q$ is $P^{text};Q^{text}$, the textual syntax
for $\exists x P$ is {\bf var} $x; P^{text}$, the textual syntax for
$\uparrow^p_{\vv{x}}P$ is {\bf proc} $p(\vv{x}); P^{text}$ {\bf end}
$p$, the textual syntax for $\downarrow^p_{\vv{x}}P$ is $P^{text};
\mbox{{\bf call }} p(\vv{x})$, and the textual syntax for $x ? P ? Q$
is {\bf if} $x$ {\bf then} $P^{text}$ {\bf else} $Q^{text}$ {\bf fi}.
Also, $\mu(u,v,P(u,v),\geq)^{text}$ is typically
$P(v,v)^{text}$.
In addition to this, of course, an $L$ program will have a syntax
specified by the language $L$.

\subsubsection{Side effects}

In a realistic system, one needs to formalize imperative
operations on data for
efficiency; for example, one may have a program that repeatedly
updates a database, or repeatedly modifies an array, graph or buffer.
Creating a new copy of a data structure each time it is modified is
inefficient.  Typically one can assign values repeatedly to a program
variable using assignment statements.  However, procedures typically
have only one definition.
In order to accommodate this distinction in $PL$,
there are both {\em procedure} and {\em data} variables, and the
semantics of a data variable $x$ in a program fragment $P$ is an
ordered pair $(\alpha,\beta)$ where $\alpha$ is the initial value of
$x$ (when $P$ begins) and $\beta$ is the final value of $x$ (when $P$
ends).  By convention, $\alpha = x^{init}$ and $\beta = x^{fin}$.  In
$PL$, a procedure $p(x,y)$ with semantics $y^{fin} = x^{init}$ can
express an assignment statement $y := x$.  There are no assignment
statements {\em per se} in $PL$, and no arithmetic or Boolean
operators.  $PL$ procedures with an appropriate specification
represent such statements and operators.  In the translation to an $L$
program, such procedures would translate to the corresponding
assignment statements and operators.

For some algorithms, such as binary tree manipulation routines,
pointer manipulations are necessary.  A proper treatment of
pointers requires a
modification to the semantics of variables.  If a variable points to
the root of a binary tree, then the semantics of the variable should
include the whole tree.  If a variable points to the root of a LISP
list, the semantics of the variable should include the entire list.
Therefore, the semantics of a variable needs to include all other
values that may be reached from the variable by a sequence of
pointers.  This implies that there are side effects.  If $x$ points to
a binary tree $T$ having $T'$ as a subtree, and $y$ points to $T'$,
then a change to a substructure of $y$ will change $x$ as well.
In general, any change to a pointer will affect any structure containing
this pointer;  this is the kind of side effect that $PL$ can formalize.

Side effects can also occur if a procedure modifies variables declared
outside the procedure body.  The syntax of $PL$-feasible languages $L$
prohibits this, to simplify reasoning about $L$ programs.  However,
read and write statements modify input and output files and buffers,
and therefore imply side effects to variables declared outside a
procedure body.  Such variables are also called {\em global variables}
for the procedure.
To handle this, $PL$-feasible languages may consider
certain {\em state variables} such as the status of input and output
files as implicit parameters of every procedure.  A procedure may also
modify global variables indirectly by side effects.
This is
difficult to detect syntactically.  We assume that this cannot happen.
There are sufficient conditions to prohibit such
side effects, such as the condition that
no pointer manipulations or array element assignments can precede
procedure definitions.

In order to reason about
the side effects of one actual parameter on another,
we assume that parameters are passed by value when possible.  For
arrays and complex data structures such as lists and binary trees,
parameters are passed by reference.  The semantics of parameters
passed by reference must include not only their value but also some
information about the address at which they are stored, in order to
determine the side effects of a change of one parameter on another.

\subsubsection{Functional and Logic Programming Languages}

Because $PL$ permits an imperative style of programming, it may be
difficult to encode abstract $PL$ programs in pure functional and
logic programming languages.  However, many functional and logic
programming languages have imperative features added for efficiency,
facilitating the translation of $PL$ programs into these languages.
Some restrictions on $PL$ programs may facilitate their translation
into pure functional and logic programming languags.  For example, if
a $PL$ program is written in a {\em single assignment} style, in which
each data variable is assigned at most once, then the translation into
functional and logic programming languages appears to be fairly
direct.  The single assignment style of programming also minimizes
side effects.  A more restrictive class of $PL$ programs are those
without any data variables, and these should be even easier to
translate into pure functional and logic programming languages.

\subsubsection{Substitutions}

There are a number of axioms and rules of inference about
substitutions in $PL$.  These are necessary in order to reason about
specific instances of general programs.  Suppose that $P(x,y)$ and
$Q(u,v)$ are program fragments in $PL(L)$.  Suppose one has assertions
$A(x,y)$ and $B(u,v)$ expressing the properties of $P$ and $Q$.  The
program fragment $P(x,y) ; Q(y,v)$ expresses a sequential composition
of $P$ and $Q$, with the variable $y$ of $Q$ replacing the variable
$u$.  It is desirable to reason about the properties of this combined
program fragment.  It is plausible to assume that the assertion
$A(x,y) \wedge B(y,v)$ would hold for the program fragment $P(x,y) ;
Q(y,v)$.  However, deriving this assumption requires axioms about how
the assertions $A$ and $B$ behave under substitutions to the programs
$P$ and $Q$.  Therefore $PL$ contains a number of axioms about
substitutions and their influence on program semantics.  These axioms
enable the derivation of properties of substitution instances of a
general program from properties of the general program, and therefore
facilitate the construction of programs in $PL$ from general building
blocks.  $PL$ restricts such reasoning to substitutions that do not
identify output variables, because this assumption simplifies the
axioms.

As an example where identifying output variables leads to unusual
behavior of instances of a general program, consider the program
$P(x,y)$ equal to $x := x+1; y := y+1$ and the assertion $A(x,y)
\equiv (x^{fin}=x^{init}+1 \wedge y^{fin}=y^{init}+1)$.  The program
$P(x,x)$ is then $x := x+1; x := x+1$ and no longer satisfies the
assertion $A(x,x) \equiv (x^{fin}=x^{init}+1 \wedge
x^{fin}=x^{init}+1)$.  Instead, $P(x,x)$ satisfies the assertion
$x^{fin}=x^{init}+2$.

Because $PL$ is a general system for reasoning about programs in
various languages $L$, it is necessary for $PL$ formulas to refer to
programs in $L$.  $PL$ views $L$ programs simply as strings in a
language, with certain program variables (names of procedures and
data variables) replaced by $PL$ variables in order to reason about
instances of general programs.

\subsection{Terminology}

$L$ is a programming language and $P$ and $Q$ denote programs or fragments
of programs in $L$.  These are sometimes written as $P^L$ and $Q^L$ to
specify $L$.  Programs in $L$ are assumed to satisfy the axioms
of the system $PL(L)$ given below.

In the notation $[\vv{x}]P(\vv{y})$, $P$ is a program fragment
containing variables $\vv{y}$ that may represent procedures or data.
Variables may appear more than once in $\vv{x}$ and $\vv{y}$.  The
variables $\vv{y}$ are the ``schema variables'' of program $P$ and
$\vv{x}$ is a listing of these variables in the order they will appear
in assertions about $P$.  The {\em free variables} $FV(P)$ of a
program $P$ are those procedures and variables of $P$ that are not
locally bound in $P$, so that they are available outside of $P$.
Other variables of $P$ are {\em bound}.  By convention all free schema
variables in $P$ must appear in $\vv{x}$.  Variables may appear in
$\vv{x}$ that do not appear free in $P$.  Such variables are also
elements of $FV(P)$, by convention.
$P(\vv{x})$ may be an
abbreviation for $[\vv{x}]P(\vv{x})$.  The {\em side effect variable}
$\psi$ is included as the last element
in the list $\vv{x}$ even though $\psi$ does not
occur in $P$; this variable $\psi$ is useful for reasoning about side
effects.  The variable $\psi$ is a data variable, and
the sort of $\psi$ is the
union of the sorts of all data variables.
The semantics of $\psi$ consists of pairs of the form $(\alpha,
\beta)$ indicating that if $\alpha$ is the value of $\psi$ at the
beginning of the execution of $P$,
then side effects of $P$ cause $\beta$ to be the value of
$\psi$ at the end of the execution of $P$.
For example, if $P$ changes a pointer at address $a$ to point to $b$,
then the semantics of the side effect variable $\psi$ for $P$ would
consist of pairs $(x_1,x_2)$ such that $x_2 = x_1$ if $x_1$ is a
structure not containing the pointer $a$, and if $x_1$ does contain
the pointer $a$, then $x_2$ would be $x_1$ with this pointer modified
to point to $b$.

The side effect variable interacts with the sequential composition
operator.  The composition $P;Q$ of two program fragments has the
precondition that $FV(P) = FV(Q)$.  This precondition enables
reasoning about side effects.  For programs without side effects, this
condition can be relaxed.  Suppose $P(x)$ has only the free data
variable $x$ and $Q(y)$ has only the free data variable $y$.  In order
to compose $P$ and $Q$, it is necessary to add $y$ as a free variable
to $P$ and $x$ as a free variable to $Q$.  The {\em new variable rule}
permits this, but requires the semantics of these new variables to
reflect the side effects of the executions of $P$ and $Q$.  Thus in
$P(x)$, the new variable $y$ has semantics reflecting the side effects
of the execution of $P$ on $y$.  Similarly, in $Q(y)$, the new
variable $x$ has semantics reflecting the side effects of the
execution of $Q$ on $x$.  Therefore in the program fragment
$P(x);Q(y)$, the overall semantics of $x$ would reflect the effect of
executing $P$, followed by the side effects of the execution of $Q$ on
$x$, and the semantis of $y$ would reflect the side effects of
executing $P$, followed by the effect of executing $Q$.

$P^{in}$ denotes the set of {\em input variables} of program $P$ and
$P^{out}$ denotes the {\em output variables}.  Input variables of $P$
are those that are free in $P$ but not defined there.  Output
variables of $P$ are variables that are defined in $P$ and may or may
not be used in $P$.
Each
variable may be {\em data}, which must be defined before it is used,
or a {\em procedure}, which can be defined after it is used.
No procedure variable may be defined twice.
$P^{proc}$ denotes the free variables of $P$ that are procedures and
$P^{data}$ denotes those that are data.  
If $p \in P^{proc}$ then $p$
takes zero or more arguments, which need not listed be among the
variables of $P$ and may either be procedures or data, and which may
be inputs or outputs of $p$.  The number of arguments of $p$ is its
{\em arity}.  The notations $p^{in}$, $p^{out}$, $p^{proc}$, and
$p^{data}$, each of which denotes a subset of $\{1, \dots, n\}$ if $n$
is the arity of $p$, indicate which arguments of $p$ are inputs,
outputs, procedures, and data.  Similarly, $p^{i,in}$ et cetera give
information about the $i^{th}$ argument of $p$, and $p^{i,j,in}$ et
cetera give information about the $j^{th}$ argument of the $i^{th}$
argument of $p$, if $p$ is a procedure.  In general, one writes
$p^{\alpha,in}$ et cetera where $\alpha$ is a sequence of integers.
To avoid dealing with sets of integers, one writes $p(\vv{x})^{in}$,
defined as $\{x_i : i \in p^{in}\}$, et cetera.  Let $x^{Type}$ be the
{\em type} of $x$, which we define as the function from $\alpha$ to
the 4-tuple $(x^{\alpha,in}, x^{\alpha,out}, x^{\alpha,data},
x^{\alpha,proc})$, for integer sequences $\alpha$.  Let
$x^{\alpha,Type}$ be the function from $\beta$ to the 4-tuple
$(x^{\alpha\beta,in}, x^{\alpha\beta,out}, x^{\alpha\beta,data},
x^{\alpha\beta,proc})$, for integer sequences $\alpha$ and $\beta$.
Thus $x^{\alpha,Type}$ is the type of $x^{\alpha}$, in a sense.  Also,
$x_P$ or $x[P]$ denotes the variable $x$ of the program $P$.

The notation $\{\vv{z} : [\vv{x}]P(\vv{y})\}$ means that $\vv{z}$ is a
sequence of values representing possible semantics of the schema
variables $\vv{x}$ that appear in $P$, $z_i$ being the semantics of
$x_i$, and $\vv{y}$ is the variables $\vv{x}$ listed in a possibly
different order.  Note that $\vv{z}$ is not a function of $P$, because
$P$ may be just a fragment of a larger program $P'$, and some of the
program variables in $P$ may be procedures that are defined elsewhere
in $P'$.
The constraint $C^L_{[\vv{x}]P}$ represents the constraint
on the semantics of $\vv{x}$ imposed by the program fragment $P^L$.
Thus $C^L_{[\vv{x}]P}(\vv{y})$ denotes that $\vv{y}$ are possible
semantics for the schema variables $\vv{x}$ that are consistent with
the program fragment $P^L$.  This is simply another notation for
$\{\vv{y} : [\vv{x}]P^L\}$.

If $x$ and $y$ are variables or terms, $x \equiv y$ means that $x$ and
$y$ are syntactically identical.  If $\vv{x}$ and $\vv{y}$ are
sequences of variables, then $\vv{x} \circ \vv{y}$ denotes the
concatenation of these two sequences.  Often this is written with a
comma as $\vv{x},\vv{y}$.  If $\vv{x}$ is the sequence $x_1, \dots,
x_n$ of variables then $\vv{\{}\vv{x}\vv{\}}$ denotes the set $\{x_1,
\dots, x_n\}$.  The notation $\vv{x} \rightarrow \vv{y}$ indicates
that $y_i \equiv y_j$ if $x_i \equiv x_j$.  A {\em variable
substitution} is a function from program variables to program
variables, often indicated by $\Theta$.  If $P$ is a program fragment
and $\Theta$ is a variable substitution, then $P\Theta$ denotes $P$
with free variables replaced as specified by $\Theta$, and bound
variables renamed to avoid captures.  A variable substitution $\Theta$
is {\em output-injective} on $P$ if for all distinct variables $x,y
\in P^{out}$, $x\Theta \not\equiv y\Theta$.  If $\Theta$ is output
injective on $P$ then $P\Theta^{out} = P^{out}\Theta$ and
$P\Theta^{in} = P^{in}\Theta - P\Theta^{out}$.  Thus a variable that
is the image of both an input and an output variable, is an output
variable.  A variable substitution may only identify variables of the
same type, both of which are either procedure variables or data
variables.

The symbol $\Theta$ typically denotes a variable substitution and
$\sigma$ typically denotes a function from integers to integers.  If
$\Theta$ is one to one it is called a {\em variable renaming} and if
$\sigma$ is also one to one it is called an {\em integer permutation}.
If $\vv{x}$ is a tuple of program variables and $\Theta$ is a variable
substitution then $\vv{x}\Theta$ denotes $x_1\Theta, \dots,
x_n\Theta$.  If $\sigma$ is an integer function and $\vv{x}$ is any
tuple, then $\vv{\sigma}(\vv{x})$ denotes $x_{\sigma(1)}, \dots,
x_{\sigma(n)}$.  Also, if $\sigma$ is an integer function and $\vv{x}$
is a tuple of program variables then $\hat{\sigma}(\vv{x})$ denotes
the variable substitution such that $\vv{\sigma}(\vv{x}) =
\vv{x}\hat{\sigma}(\vv{x})$, that is, the substitution $\{x_1
\rightarrow x_{\sigma(1)}, \dots, x_n \rightarrow x_{\sigma(n)}\}$.
The side effect variable $\psi$ is always the last element of
the list $\vv{x}$ of variables, which means that no variable substitution
or integer function can change this property.  For example,
for all variable substitutions $\Theta$, $\psi\Theta = \psi$.

The symbols $f$ and $g$ typically denote functions from variables to
their semantics.  If $f$ is a function on free variables of $P$ then
$\vv{f}(\vv{x})$ denotes $(f(x_1), \dots, f(x_n))$.

The formula $\Sigma x A[x]$ means $A[x] \wedge A[y] \supset x = y$,
that is, $A$ is {\em exclusive} for $x$.

The axioms of $PL(L)$ are as follows:

\subsection{Definitions}
\paragraph{Definition of constraint $C$ in terms of colon notation}

\begin{equation}
C^L_{[\vv{x}]P}(\vv{y}) \equiv \{\vv{y} : [\vv{x}]P^L\}
\end{equation}

\paragraph{Definition of $R$ on programs}

If $R$ is a relation on semantics of program variables then

\begin{equation}
R^L([\vv{x}]P) \equiv \forall \vv{y}(\{\vv{y} : [\vv{x}]P^L\} \supset
R(\vv{y}))
\label{eq:progdef}
\end{equation}

$R^{L,\psi}$ explicitly considers the side effect variable $\psi$.
$R^{L,\vv{\psi}}$ does not.  If neither superscript appears, either
meaning is possible.

\subsection{Axioms about $P^{in}$ and $P^{out}$}

The general idea is that if a variable is an input variable in one
part of a program and an output variable elsewhere, it is an output
variable for the whole program.

If $\Theta$ is an output injective variable substitution then

\begin{equation}
(P(\vv{x})\Theta)^{in} = P(\vv{x})^{in}\Theta - P(\vv{x})^{out}\Theta
\label{eq:isubst}
\end{equation}
If $\Theta$ is an output injective variable substitution then

\begin{equation}
(P(\vv{x})\Theta)^{out} = P(\vv{x})^{out}\Theta
\label{eq:osubst}
\end{equation}

\begin{equation}
(P(\vv{x});Q(\vv{y}))^{in} = P(\vv{x})^{in} \cup Q(\vv{y})^{in}
-(P(\vv{x})^{out} \cup Q(\vv{y})^{out})
\label{eq:icomp}
\end{equation}

\begin{equation}
(P(\vv{x});Q(\vv{y}))^{out} = P(\vv{x})^{out} \cup Q(\vv{y})^{out}
\label{eq:ocomp}
\end{equation}

\begin{equation}
(u ? P(\vv{x}) ? Q(\vv{y}))^{in} = P(\vv{x})^{in} \cup Q(\vv{y})^{in}
\cup \{u\} -(P(\vv{x})^{out} \cup Q(\vv{y})^{out})
\label{eq:iif}
\end{equation}

\begin{equation}
(u ? P(\vv{x}) ? Q(\vv{y}))^{out} = P(\vv{x})^{out} \cup Q(\vv{y})^{out}
\label{eq:oif}
\end{equation}

\begin{equation}
\uparrow^p_{\vv{y}}P(\vv{x},\vv{y})^{in} = P(\vv{x},\vv{y})^{in} -
\vv{\{}\vv{y}\vv{\}}
\label{eq:iuparrow}
\end{equation}

\begin{equation}
\uparrow^p_{\vv{y}}P(\vv{x},\vv{y})^{out}=P(\vv{x},\vv{y})^{out} \cup \{p\} -
\vv{\{}\vv{y}\vv{\}}
\label{eq:ouparrow}
\end{equation}

\begin{equation}
\downarrow^p_{\vv{y}}P(\vv{x},p)^{in} = P(\vv{x},p)^{in} \cup p(\vv{y})^{in}
- \downarrow^p_{\vv{y}}P(\vv{x},p)^{out}
\label{eq:idownarrow}
\end{equation}

\begin{equation}
\downarrow^p_{\vv{y}}P(\vv{x},p)^{out}=P(\vv{x},p)^{out} \cup
p(\vv{y})^{out}
\label{eq:odownarrow}
\end{equation}

\begin{equation}
\exists wP^{in} = P^{in}
\label{eq:idelete}
\end{equation}
\begin{equation}
\exists wP^{out} = P^{out}- \{w\}
\label{eq:odelete}
\end{equation}
\begin{equation}
\mu(u,v,P,\geq)^{in}=P^{in}-\{u\}
\label{eq:ifixpoint}
\end{equation}
\begin{equation}
\mu(u,v,P,\geq)^{out}=P^{out}
\label{eq:ofixpoint}
\end{equation}

\subsection{Axioms about free variables}

\begin{equation}
FV(P) = P^{in} \cup P^{out}
\end{equation}

\subsubsection{Consequences of this axiom}

\begin{equation}
FV(P;Q)=FV(P)\cup FV(Q)
\end{equation}

\begin{equation}
FV(x ? P ? Q)=FV(P)\cup FV(Q) \cup \{x\}
\end{equation}

\begin{equation}
FV(\uparrow^p_{\vv{y}}P(\vv{x},\vv{y}))=\{p\}\cup \vv{\{}\vv{x}\vv{\}}
\end{equation}

\begin{equation}
FV(\downarrow^p_{\vv{y}}P(\vv{x},p))=\vv{\{}\vv{x}\vv{\}} \cup
\vv{\{}\vv{y}\vv{\}} \cup \{p\}
\end{equation}

\begin{equation}
FV(\exists zP) = FV(P)-\{z\}
\end{equation}

\begin{equation}
FV(\mu(u,v,P(\vv{x},u,v),\geq)) = \vv{\{}\vv{x}\vv{\}} \cup \{v\}
\end{equation}

\subsection{Axioms about arguments to procedure variables}

If $x \in FV(P)$ then
\begin{equation}
x^{Type}[P;Q]=x^{Type}[P]
\end{equation}

If $x \in FV(Q)$ then
\begin{equation}
x^{Type}[P;Q]=x^{Type}[Q]
\end{equation}

\begin{equation}
p^{i\alpha,Type}[\uparrow^p_{\vv{y}}P]=y_i^{\alpha,Type}[P]
\end{equation}

If $x \in FV(\uparrow^p_{\vv{y}}P) - \{p\}$ then
\begin{equation}
x^{Type}[\uparrow^p_{\vv{y}}P]=x^{Type}[P]
\end{equation}

\begin{equation}
y_i^{\alpha,Type}[\downarrow^p_{\vv{y}}P]=p^{i\alpha,Type}[P]
\end{equation}

If $x \in FV(\downarrow^p_{\vv{y}}P) - \vv{\{}\vv{y}\vv{\}}$ then
\begin{equation}
x^{Type}[\downarrow^p_{\vv{y}}P]=x^{Type}[P]
\end{equation}

If $x \in FV(\exists y P) - \{y\}$ then
\begin{equation}
x^{Type}[\exists y P]=x^{Type}[P]
\end{equation}

If $z \in FV(\mu(u,v,P,\geq)(\vv{x},\vv{y},v)) - \{y\}$ then
\begin{equation}
z^{Type}[\mu(u,v,P,\geq)(\vv{x},\vv{y},v)]=z^{Type}[P]
\end{equation}

For any substitution $\Theta$,
\begin{equation}
x\Theta^{Type}[P\Theta] = x^{Type}[P]
\end{equation}

and
\begin{equation}
P\Theta^{proc} = P^{proc}\Theta
\end{equation}

and
\begin{equation}
P\Theta^{data} = P^{data}\Theta
\end{equation}

\subsubsection{Consequences of the above}

\begin{equation}
p^{in}(\vv{y})[\uparrow^p_{\vv{y}}P]=P^{in} \cap \vv{\{}\vv{y}\vv{\}}
\end{equation}

\begin{equation}
p^{out}(\vv{y})[\uparrow^p_{\vv{y}}P]=P^{out} \cap \vv{\{}\vv{y}\vv{\}}
\end{equation}

\subsection{Axioms about program forming operations}
\label{PL.axioms}

\paragraph{Preconditions for the sequential composition operator}

The operation $P;Q$ is allowed if no procedure variable $x$ is in
$P^{out} \cap Q^{out}$
and if $P^{data} = Q^{data}$.  The latter condition can be satisfied
by adding extra variables to $P$ and $Q$ if necessary using the new
variable axiom that appears below.

\paragraph{Definition of $\phi$ operator}
The definition of the sequential composition operator makes use of the
$\phi$ operator, defined as follows:
If $f$ and $g$ are semantic functions and $P$ and $Q$ are program
fragments then the semantic function $\phi_{P,Q}(f,g)$ satisfies
the following:
\begin{enumerate}
\item if $z \in P^{proc}$ then $\phi_{P,Q}(f,g)(z) = f(z)$.
\item if $z \in Q^{proc} - P^{proc}$ then $\phi_{P,Q}(f,g)(z) = g(z)$.
\item if $z \in P^{data}\cap Q^{data}$ and
$f(z) = (\alpha,\beta)$ and $g(z) = (\beta,\gamma)$ then
$\phi_{P,Q}(f,g)(z) = (\alpha,\gamma)$.
\end{enumerate}
\noindent
If $f$ and $g$ give a semantics for $P$ and $Q$, then $\phi(f,g)$ gives
a semantics for $P;Q$.

\paragraph{Preconditions for $\phi(f,g)$ operator}
$prec(\phi,P,Q,f,g)$ specifies
\begin{enumerate}
\item if $z \in P^{proc} \cap Q^{proc}$ then $f(z) = g(z)$.
\item if $z \in P^{data} \cap Q^{data}$ then $\exists \alpha \beta \gamma
(f(z)=(\alpha,\beta) \wedge g(z) = (\beta,\gamma))$.
\end{enumerate}

\paragraph{Sequential composition axiom}

\begin{equation}
\{\vv{h}(\vv{x}\circ\vv{y}):[\vv{x},\vv{y}]P(\vv{x});Q(\vv{y})\} \equiv
\exists f g (\{\vv{f}(\vv{x}):[\vv{x}]P(\vv{x})\} \wedge
 \{\vv{g}(\vv{y}):[\vv{y}]Q(\vv{y})\}) \wedge h = \phi_{P,Q}(f,g) \wedge
prec(\phi,P,Q,f,g).
\label{eq:composition}
\end{equation}

\paragraph{Conditional axiom}

\begin{eqnarray}
\{w,\vv{u},\vv{v}:[z,\vv{x},\vv{y}]z ? P(\vv{x}) ? Q(\vv{y})\} & \equiv &
(w={\bf true} \wedge \{\vv{u}:[\vv{x}]P(\vv{x})\}) \vee
 (w={\bf false} \wedge \{\vv{v}:[\vv{y}]Q(\vv{y})\}) \nonumber \\
&&\mbox{ if } \vv{x} = FV(P) \wedge \vv{y} = FV(Q)
\label{eq:if}
\end{eqnarray}

Here $P$ or $Q$ may be empty.

\paragraph{Deleting output axiom}

Intuitively, this axiom declares $z$ to be a local variable.

\begin{equation}
\forall \vv{x}'\vv{y}'(\exists z'\{\vv{x}',z',\vv{y}' :
[\vv{x},z,\vv{y}]P\} \equiv \{\vv{x}',\vv{y}' : [\vv{x},\vv{y}]\exists zP\})
\label{eq:deleteout2}
\end{equation}

\paragraph{Preconditions for procedure operator axiom}

The operation $\uparrow^p_{\vv{y}} P(\vv{x},\vv{y})$ is allowed if
no variable $x_i$ is in $P^{out}$, and if no execution of $p$ can have
side effects on the variables $\vv{x}$.
That is, $R^L(P(\vv{x},\vv{y}))$ where $R(\vv{u},\vv{v}) \equiv
\forall i (u_i^{init}=u_i^{fin})$.

\paragraph{Procedure operator axioms}

Intuitively, these axioms define a new procedure $p$ having the
formal parameters $\vv{y}$.

\begin{equation}
\forall \vv{u} \exists q \{\vv{u},q : [\vv{x},p] \uparrow^p_{\vv{y}}
P(\vv{x},\vv{y})\}
\label{eq:procedure1}
\end{equation}

\begin{equation}
\forall p q \vv{u} (\{\vv{u},q : [\vv{x},p] \uparrow^p_{\vv{y}}
P(\vv{x},\vv{y})\} \supset \forall \vv{v} (\{\vv{u},\vv{v} :
[\vv{x},\vv{y}]P(\vv{x},\vv{y})\} \equiv q(\vv{v})))
\label{eq:procedure2}
\end{equation}
where $p$ is a new variable or an input variable of $P$.

\paragraph{Preconditions for application axiom}

The operation $\downarrow^p_{\vv{y}}P$ is allowed if
$P; \downarrow^p_{\vv{y}}$ is allowed.
The fragment $P$ may be empty, in which case there are no
preconditions and $\downarrow^p_{\vv{y}}P$ is equivalent to
$\downarrow^p_{\vv{y}}$.

\paragraph{Application axioms}

Intuitively, this operator calls a procedure $p$ with actual
parameters $\vv{y}$.

\begin{equation}
\forall \vv{u} \vv{v} q (\{\vv{u},\vv{v},q : [\vv{x},\vv{y},p]
\downarrow^p_{\vv{y}}P(\vv{x},p)\}
\equiv \{\vv{u},\vv{v},q : [\vv{x},\vv{y},p]
(P(\vv{x},p);\downarrow^p_{\vv{y}})\})
\end{equation}

\begin{equation}
\forall \vv{v}q(\{\vv{v},q : [\vv{y},p]\downarrow^p_{\vv{y}}\}
\equiv q(\vv{v}))
\end{equation}

\paragraph{Least fixpoint axiom}
If
\[\forall \vv{x}' u' \Sigma \vv{y}' v' \{\vv{x}',u',\vv{y},v' :
P(\vv{x},u,\vv{y},v)\}\]
and $\vv{y},v$ are outputs and $\vv{x},u$ are inputs of
$P$ and $u,v$ are procedure variables then
\begin{eqnarray}
\{\vv{x}',\vv{y}',w' : \mu(u,v,P,\geq)(\vv{x},\vv{y},v)\} & \equiv &
\nonumber \\ (\{\vv{x}', w', \vv{y}', w' : P(\vv{x}, v, \vv{y}, v)\} &
\wedge & \forall z' \forall \vv{y}'' (\{\vv{x}', z', \vv{y}'', z' :
P(\vv{x}, u, \vv{y}, v)\} \supset z' \geq w'))
\label{eq:leastfixpoint}
\end{eqnarray}

\subsection{Axioms about variables}

\paragraph{Correspondence axiom}

This axiom implies that semantics for procedure variables not
appearing free in $P$ can be arbitrary.  However, data variables, even
not free in $P$, can be influenced by side effects of the execution of
$P$.  Note that $([\vv{x}]P)^{data}$ may include data variables in
$\vv{x}$ that are not free in $P$.

If $\{(u_i,x_i) : x_i \in FV(P) \cup ([\vv{x}]P)^{data}\} =
\{(v_j,y_j) : y_j \in FV(P) \cup ([\vv{y}]P)^{data}\}$
then
\begin{equation}
\{\vv{u} : [\vv{x}]P\} \supset\{\vv{v} : [\vv{y}]P\}.
\label{eq:corresp}
\end{equation}

\subparagraph{Alternate version}
If $\forall u \in FV(P) \cup ([\vv{x}]P)^{data}\cup ([\vv{y}]P)^{data}
f(u)=g(u)$ then
\begin{equation}
\{\vv{f}(\vv{x}) : [\vv{x}]P\} \supset\{\vv{g}(\vv{y}) : [\vv{y}]P\}.
\label{eq:corresp2}
\end{equation}

\paragraph{New variable axiom}
If $y$ is a data variable that does not appear free in $P$ 
or in $\vv{x}$ then
\begin{equation}
\{\vv{u},v,w : [\vv{x},y,\psi]P\} \equiv
\{\vv{u},v : [\vv{x},\psi]P\} \wedge
\{\vv{u},w : [\vv{x},\psi]P\}
\end{equation}

\paragraph{Variable renaming axiom}

If $\Theta$ is a variable renaming then
\begin{equation}
\{\vv{y} : [\vv{x}]P(\vv{x})\} \supset
\{\vv{y} : [\vv{x}\Theta]P(\vv{x}\Theta)\}
\label{eq:variablerename}
\end{equation}
Note that variable renamings are variable substitutions and are
output injective.

\subparagraph{Special case}

If $\sigma$ is an integer permutation then
\begin{equation}
\{\vv{y} : [\vv{x}]P(\vv{x})\} \supset
\{\vv{y} : [\vv{\sigma}(\vv{x})]P(\vv{\sigma}(\vv{x}))\}
\label{eq:variablerename2}
\end{equation}

\paragraph{Equality axiom}

\begin{equation}
(\{\vv{y} : [\vv{x}]P\} \wedge x_i \equiv x_j) \supset y_i = y_j
\label{eq:equality}
\end{equation}

\paragraph{Definitional independence axiom}

The idea of this axiom is that any semantics for an instance of a program
must also satisfy the constraints of the general program, that is, the
definitions are independent of the instance of the program.

If $\vv{x} \rightarrow \vv{y}$ and $P(\vv{y})$ does not identify distinct
outputs or data variables of $P(\vv{x})$ then
\begin{equation}
\forall \vv{z} (\{\vv{z} : [\vv{y}] P(\vv{y})\} \supset 
\{\vv{z} : [\vv{x}] P(\vv{x})\})
\label{eq:defind1}
\end{equation}

\subparagraph{Alternative version}

If $\Theta$ is a variable substitution that is output-injective on $P$
and does not identify two data variables then

\begin{equation}
\forall \vv{z} (\{\vv{z} : [\vv{x}\Theta] P(\vv{x}\Theta)\} \supset 
\{\vv{z} : [\vv{x}] P(\vv{x})\})
\label{eq:defind2}
\end{equation}

\subparagraph{Examples of definitional independence}

Let $P(x,y,u,v)$ be the program $y := x+1; v := u+1$.  Consider the
program $P(x,y,y,v)$ which is $y := x+1; v := y+1$.  Then
$\{((1,1),(0,2),(0,2),(0,3)) : P(x,y,y,v)\}$.  Definitional
independence asserts that $\{((1,1),(0,2),(0,2),(0,3)): P(x,y,u,v)\}$
which is not correct because $u$ is not modified in $P(x,y,u,v)$.  The
problem is that two data variables have been identified.
However, there is a semantics for $P(x,y,u,v)$ in which the final values
of the variables $(x,y,u,v)$ are $(1,2,2,3)$, respectively.

Definitional independence applies to recursive procedure definitions.  For this
example, assume that $L$ has function procedures.  Let
$P(f,g,h,k)$ define $f(x)$ as ``if $x = 0$ then $0$ else $g(x) +
k(h(x))$.''  Then $P(f,g,h,f)$ defines $f(x)$ as ``if $x = 0$ then $0$
else $g(x) + f(h(x))$.''  Any semantics for $P(f,g,h,f)$ is also a
semantics for $P(f,g,h,k)$.

\subsection{Consequences of the above axioms}

\paragraph{Permutation axiom}

If $\sigma$ is an integer permutation then
\begin{equation}
\{\vv{y} : [\vv{x}]P\} \supset \{\vv{\sigma}(\vv{y}) :
[\vv{\sigma}(\vv{x})]P\}
\label{eq:perm}
\end{equation}

\paragraph{Variable independence axiom}
If $x_i$ is a procedure variable that does not appear free in $P$ then
$\{\vv{y} : [\vv{x}]P\}$ does not depend on $y_i$, so
\begin{equation}
\{\vv{y} : [\vv{x}]P\}\supset \forall y_i\{\vv{y}
: [\vv{x}]P\}.
\label{eq:varind}
\end{equation}

\section{Definitional Independence and Fixed Points}

In order to justify the reasonableness of the preceding axioms,
it is possible to show that programming languages with certain
properties satisfy definitional independence and the fixpoint
axiom.  For the former,
suppose that $\vv{y}$ and $\vv{z}$ are procedure variables of $P$
and $\vv{w}$ are data variables.  Let $\vv{w}^{init}$ be
$(w_1^{init}, \dots, w_n^{init})$ and let
$\vv{w}^{fin}$ be
$(w_1^{fin}, \dots, w_n^{fin})$.
Writing the assertion $\{(\vv{u},\vv{v},\vv{x}):P(\vv{y},\vv{z},\vv{w})\}$ as
$\{[\vv{u},\vv{x}^{init} \rightarrow \vv{v},\vv{x}^{fin}]:
P[\vv{y},\vv{w}^{init} \rightarrow\vv{z},\vv{w}^{fin}]\}$
indicates that $\vv{y}$ are the input variables of $P$, $\vv{z}$ are
the output variables, $\vv{u}$ are possible semantics of $\vv{y}$,
$\vv{v}$ are possible semantics of $\vv{z}$, and $\vv{x}$ are possible
semantics of $\vv{w}$.  For simplicity, $\vv{w}$ is ignored from now on,
because it does not affect the argument.

\begin{Define}
\label{def:denotational}
A programming language $L$ is {\em denotational} if for all
programs $P$ in $L$ there is a monotonic and continuous
functional $\tau_P$ such that $\{[\vv{u} \rightarrow \vv{v}]:P[\vv{y}
\rightarrow\vv{z}]\}$ iff $\vv{v} = \tau_P(\vv{u})$ and
$(\vv{y},\vv{z}) \rightarrow (\vv{u},\vv{v})$.  (Here it is assumed
that $\vv{y}$ and $\vv{z}$ are disjoint.)  Also, if $\Theta$ is an
output injective variable substitution, then $\{[\vv{u}' \rightarrow
\vv{v}']:P[\vv{y}\Theta \rightarrow\vv{z}\Theta]\}$ iff $(\vv{u}',
\vv{v}')$ is the minimal element of the set $\{(\vv{u}'',\vv{v}'') :
\vv{v}'' = \tau_P(\vv{u}'') \mbox{ and } (\vv{y}\Theta,\vv{z}\Theta)
\rightarrow (\vv{u}'',\vv{v}'') \mbox{ and } u_i'' = u_i' \mbox{ if }
y_i\Theta \not\in \vv{z}\Theta\}$.
\end{Define}

The idea is that if $\Theta$
identifies input and output variables of $P$, then these variables are
considered as output variables and they need to be minimized subject
to the equation $\vv{v}'' = \tau_P(\vv{u}'')$.  The condition
$(\vv{y},\vv{z}) \rightarrow (\vv{u},\vv{v})$ means that if two elements
$y_i$ and $y_j$ are the same, their semantics must be the same, and
similarly for elements of $\vv{z}$ and for common elements of
$\vv{y}$ and $\vv{z}$.

This condition is reasonable; it states that any inputs to a partial
program have definitions outside the partial program,
so nothing can be assumed about their
semantics.  But any procedure that is an output of the partial program
has a definition in the partial program,
and therefore
has the denotationally smallest semantics that satisfies the
definition.  For example, in the partial program $P(x,y \rightarrow
z,w)$, the definitions of
$x$ and $y$ occur outside of $P$ but the definitions of $z$ and $w$ occur
in $P$, so
the semantics of $z$ and
$w$ are constrained by the semantics of $x$ and $y$ and the
definitions of $z$ and $w$ in terms of $x$ and $y$.  Now consider
$P(x,y \rightarrow y,w)$.  This is an instance of $P(x,y \rightarrow
z, w)$.  The procedure $y$ now has a recursive definition, and
receives the least possible semantics satisfying its definition.  The
definition of the procedure $x$ occurs elsewhere, so that the
semantics of $x$ is arbitrary.  But the semantics of $x$ determines the
semantics of $y$ and $w$.

\begin{theorem}  If $L$ is denotational then every program $P$ in
$L$ satisfies definitional independence.
\end{theorem}

\begin{proof}
Suppose $\{[\vv{u}' \rightarrow \vv{v}']:P[\vv{y}\Theta
\rightarrow\vv{z}\Theta]\}$.  First assume that $\vv{y}\Theta$ and
$\vv{z}\Theta$ are disjoint.  This means that no element of
$\vv{y}\Theta$ is in $\vv{z}\Theta$.
Since $L$ is denotational,
it must be that $(\vv{y}\Theta, \vv{z}\Theta) \rightarrow
(\vv{u}', \vv{v}')$ and
$\vv{v}' =
\tau_P(\vv{u}')$ by definition \ref{def:denotational}.
Since $(\vv{y},\vv{z}) \rightarrow (\vv{y}\Theta, \vv{z}\Theta)$
and $(\vv{y}\Theta, \vv{z}\Theta) \rightarrow
(\vv{u}', \vv{v}')$,  $(\vv{y},\vv{z}) \rightarrow (\vv{u}', \vv{v}')$ as well.
Because  $\vv{v}' = \tau_P(\vv{u}')$ and
$(\vv{y},\vv{z}) \rightarrow (\vv{u}', \vv{v}')$,
$\{(\vv{u}',\vv{v}') : P(\vv{y},\vv{z})\}$ also.

Now consider the case when $\vv{y}\Theta$ and $\vv{z}\Theta$ are not
disjoint.  Write $(\vv{y}\Theta, \vv{z}\Theta)$ as $(\vv{y}'\Theta,
\vv{x}, \vv{x}, \vv{z}'\Theta)$, indicating by $\vv{x}$ the parts of
$\vv{y}$ and $\vv{z}$ that are identified by $\Theta$ and by $\vv{y}'$
and $\vv{z}'$ the remaining parts of $\vv{y}$ and $\vv{z}$.
Similarly, write $(\vv{u}', \vv{v}')$ as
$(\vv{u}'',\vv{w},\vv{w},\vv{v}'')$.  The idea of the definition is
that $\vv{w}$ and $\vv{v}$ are chosen to be as small as possible
subject to the condition that $(\vv{w},\vv{v}'') =
\tau_P(\vv{u}'',\vv{w})$, but $\vv{u}''$ is chosen to be equal to the
corresponding components of $\vv{u}'$, which are not constrained.

Since $L$ is denotational and $\{[\vv{u}' \rightarrow
\vv{v}']:P[\vv{y}\Theta \rightarrow\vv{z}\Theta]\}$,  $(\vv{u}',
\vv{v}')$ is the minimal element of the set $\{(\alpha,\beta) :
\beta = \tau_P(\alpha) \mbox{ and } (\vv{y}\Theta,\vv{z}\Theta)
\rightarrow (\alpha,\beta) \mbox{ and } \alpha_i = u_i' \mbox{ if }
y_i\Theta \not\in \vv{z}\Theta\}$.

From this it follows that $\vv{v}' = \tau_P(\vv{u}')$ and
$(\vv{y}\Theta,\vv{z}\Theta)
\rightarrow (\vv{u}',\vv{v}')$.

We need to show that $\{[\vv{u}' \rightarrow \vv{v}']:P[\vv{y}
\rightarrow\vv{z}]\}$, that is,
$(\vv{u}',\vv{v}')$ is the minimal element of the set $\{(\alpha,\beta) :
\beta = \tau_P(\alpha) \mbox{ and } (\vv{y},\vv{z})
\rightarrow (\alpha,\beta) \mbox{ and } \alpha_i = u_i' \mbox{ for all } i\}$.

First, $\vv{v}' = \tau_P(\vv{u}')$ as noted above.

Second, $(\vv{y},\vv{z}) \rightarrow (\vv{u}',\vv{v}')$ because
$(\vv{y},\vv{z}) \rightarrow (\vv{y}\Theta,\vv{z}\Theta)$ for
any $\Theta$ and (as noted above)
$(\vv{y}\Theta,\vv{z}\Theta) \rightarrow
(\vv{u}',\vv{v}')$.

Finally, we need to show that if
$\beta = \tau_P(\alpha) \mbox{ and } (\vv{y},\vv{z})
\rightarrow (\alpha,\beta) \mbox{ and } \alpha_i = u_i' \mbox{ for all } i$
then $\alpha \ge \vv{u}'$ and $\beta \ge \vv{v}'$.
But if $\alpha_i = u_i' \mbox{ for all } i$ then $\alpha = \vv{u}'$.
Thus $\beta = \tau_P(\alpha) = \tau_P(\vv{u}') = \vv{v}'$.
Thus $\alpha = \vv{u}'$ and $\beta = \vv{v}'$.
\end{proof}

As an example where the theorem fails if $\Theta$ identifies two
data variables, consider the program $[x,y]x := x+1$ and its instance
$[x,x]x := x+1$.  The latter has the semantics $((0,1),(0,1))$ but
not the former, because the value of $y$ may not change.

Because functional, logic, and procedural languages are denotational, with
reasonable definitions of their semantics,
it is reasonable to assume that all these languages also satisfy
definitional independence, and that all the inference rules in $PL$
apply to all such languages.

In practice, one may use $PL$ without a formal proof that the
languages $L$ satisfy definitional independence, to obtain programs
that may have added reliability even if there is no formal proof of
correctness.

The denotational property also suffices to justify the least fixpoint
axiom.

\begin{theorem}
Suppose $L$ is denotational.  Then
the least fixpoint axiom is satisfied if one lets
$\mu(u,v,P,\geq)$ be $P\Theta$ where $\Theta$ maps $u$ to $v$ but
leaves all other variables unchanged.
\end{theorem}

\begin{proof}

We show the least
fixpoint axiom, axiom
\ref{eq:leastfixpoint}, which is the following:
If
\[\forall \vv{x}' u' \Sigma \vv{y}' v' \{\vv{x}',u',\vv{y},v' :
P(\vv{x},u,\vv{y},v)\}\]
and $\vv{y},v$ are outputs and $\vv{x},u$ are inputs of
$P$ then
\begin{eqnarray}
\{\vv{x}',\vv{y}',w' : \mu(u,v,P,\geq)(\vv{x},\vv{y},v)\} & \equiv &
\nonumber \\ (\{\vv{x}', w', \vv{y}', w' : P(\vv{x}, v, \vv{y}, v)\} &
\wedge & \forall z' \forall \vv{y}'' (\{\vv{x}', z', \vv{y}'', z' :
P(\vv{x}, u, \vv{y}, v)\} \supset z' \geq w'))
\end{eqnarray}

The hypothesis $\forall \vv{x}' u' \Sigma \vv{y}' v'
\{\vv{x}',u',\vv{y},v' : P(\vv{x},u,\vv{y},v)\}$ is satisfied because
$L$ is denotational.  Suppose $\{\vv{x}',\vv{y}',w' :
\mu(u,v,P,\geq)(\vv{x},\vv{y},v)\}$.  Defining $\mu$ as in the
theorem, and using the correspondence axiom, this is equivalent to
$\{\vv{x}',w',\vv{y}',w' : P(\vv{x},v,\vv{y},v)\}$.  For the remaining
part, write $\{\vv{x}',w',\vv{y}',w' : P(\vv{x},v,\vv{y},v)\}$ as
$\{[\vv{x}',w' \rightarrow \vv{y}',w']:P[\vv{x},v \rightarrow
\vv{y},v]\}$ and write $\{\vv{x}',z',\vv{y}'',z' :
P(\vv{x},u,\vv{y},v)\}$ as $\{[\vv{x}',z' \rightarrow
\vv{y}'',z']:P[\vv{x},u \rightarrow \vv{y},v]\}$.
Because $L$ is
denotational, $\{[\vv{x}',z' \rightarrow \vv{y}'',z']:P[\vv{x},u
\rightarrow \vv{y},v]\}$ iff $(\vv{x}',z',\vv{y}'',z')$ is the
minimal element of the set $\{(\vv{x}'',z'',\vv{y}''',z'') :
(\vv{x}'',z'') = \tau_P(\vv{y}''',z'') \mbox{ and }
((\vv{x},u),(\vv{y},v)) \rightarrow
((\vv{x}'',z''),(\vv{y}''',z'')) \mbox{ and } \vv{x}'' = \vv{x}'$ and
$z'' = z'$
(because $\vv{x}$ and $\vv{y}$ are assumed disjoint).
The condition $((\vv{x},u),(\vv{y},v)) \rightarrow
((\vv{x}'',z''),(\vv{y}''',z''))$ is true because $\vv{x},u,\vv{y}$,
and $v$ are pairwise disjoint.
Thus the only constraint on $(\vv{x}',z',\vv{y}'',z')$ is that
$(\vv{x}',z') = \tau_P(\vv{y}'',z')$.  However, because $\Theta$ identifies
$u$ and $v$, the corresponding constraint on $(\vv{x}',w',\vv{y}')$ is that
$w'$ should be minimal satisfying $(\vv{x}',w') = \tau_P(\vv{y}',w')$.
Therefore $z' \geq w'$ as specified above.
The other direction follows by
similar reasoning, because $L$ is denotational.

\end{proof}

\section{Inference Rules}
\label{relational.rules}
Section \ref{PL.axioms}
contains axioms for program language semantics expressed
in terms of the operator $:$.  These axioms lead to {\em
relational} inference rules
for deriving assertions $R^L(P)$ where $P$ is an $L$ program and $R$ is a
relation on semantics of program variables, using the definition

\[R^L([\vv{x}]P) \equiv \forall \vv{y}(\{\vv{y} : [\vv{x}]P\} \supset
 R(\vv{y}))\]

If $R$ is such a relation and $\sigma$ is an integer function then
$R\sigma$ denotes the relation such that $R\sigma(\vv{y}) \equiv
R(\vv{\sigma}(\vv{y}))$, that is, $R\sigma(y_1, \dots, y_n)$ iff
$R(y_{\sigma(1)}, \dots, y_{\sigma(n)})$.

The system {\em relational $PL(L)$} consists of the following inference
rules, which are
consequences of the axioms given in
section \ref{PL.axioms}:

\RRule{R_1^L(P), R_1 \supset R_2}{R_2^L(P)}{\mbox{ Underlying logic rule}}

\RRule{R_k^L(P), k \mbox{ arbitrary }}
{\forall k R_k^L(P)}{\mbox{ Universal quantification rule}}

\paragraph{Rules about variables}

\RRule{R^L([\vv{x}]P(\vv{x})), \Theta \mbox{ is a variable renaming}}
{R^L([\vv{x}\Theta]P(\vv{x}\Theta))}
{\mbox{Variable renaming rule}}

\RRule{R^{L,\vv{\psi}}([\vv{x}]P(\vv{x})), \sigma \mbox{ is an integer permutation}}
{R\sigma^{L,\vv{\psi}}([\vv{\sigma}^{-1}(\vv{x})]P(\vv{x}))}
{\mbox{Permutation rule 1}}

\RRule{R^L([\vv{x}]P(\vv{x})), \sigma \mbox{ is an integer permutation}}
{R\sigma^L([\vv{x}]P(\vv{\sigma}(\vv{x})))}
{\mbox{Permutation rule 2}}

\RRule{R^L(P(\vv{x})), x_i \equiv x_j, R_1(\vv{y}) \equiv (R(\vv{y}) \wedge y_i=y_j)}
{R_1^L(P(\vv{x}))}{\mbox{Equality rule}}

\RRule{R^L([\vv{x}]P(\vv{x})),\sigma \mbox{ is an integer function such that }
\vv{\sigma} \mbox{ does not identify data variables of } \vv{x}}
{R\sigma^L([\vv{x}]P(\vv{\sigma}(\vv{x}))}{\mbox{Substitution rule}}

\RRule{R^L([\vv{x}]P), \forall f,g(
R(\vv{f}(\vv{x})) \wedge \forall u \in FV(P) \cup ([\vv{x}]P)^{data}
(f(u)=g(u))) \supset R_1(\vv{g}(\vv{y}))}
{R_1^L([\vv{y}]P)}{\mbox{Correspondence rule}}

\RRule{R^L([\vv{x}]P(\vv{x})), \Theta \mbox{ output injective on } P
\mbox{ and does not identify data variables of } P}
{R^L([\vv{x}\Theta]P(\vv{x}\Theta))}{\mbox{Definitional independence rule}}

\RRule{R^{L,\psi}L([\vv{x},\psi]P), y \mbox{ a new data variable},
R(\vv{u},v) \wedge R(\vv{u},w) \supset R_1(\vv{u},v,w)}
{R_1^{L,\psi}([\vv{x},y,\psi]P)}{\mbox{New variable rule}}

\paragraph{Rules about program operations}

\RRule{R_1^L(P_1(\vv{y})), R_2^L(P_2(\vv{z})), 
prec(\phi,P,Q,f,g),
\forall \vv{f} \vv{g}(R_1(\vv{f}(\vv{y})) \wedge R_2(\vv{g}(\vv{z})) \supset
R(\phi_{P,Q}(\vv{f},\vv{g})(\vv{y} \circ \vv{z})))}
{R^L([\vv{y},\vv{z}]P_1(\vv{y}); P_2(\vv{z}))}{\mbox{Composition rule}}

\RRule{
\begin{array}{c}
R_1^L(P_1(\vv{y})), R_2^L(P_2(\vv{z})), 
\vv{y} \subseteq FV(P_1), \vv{z} \subseteq FV(P_2),\\
\forall x' \vv{y}' \vv{z}'(x' \wedge R_1(\vv{y}')) \vee 
(\neg x' \wedge R_2(\vv{z}')) \supset
R(x',\vv{y}',\vv{z}')
\end{array}}
{R^L([x,\vv{y},\vv{z}]x ? P_1(\vv{y}) ? P_2(\vv{z}))}{\mbox{Conditional rule}}

\RRule{R^L([\vv{u},x,\vv{v}]P),
R_1(\vv{u},\vv{v}) \equiv \exists x R(\vv{u},x,\vv{v})}
{R_1^L([\vv{u},\vv{v}]\exists xP)}{\mbox{Output deletion rule}}

\RRule{R^L(P(\vv{x},\vv{y})), R_1(\vv{u},q) \equiv \forall \vv{v}
(q(\vv{v}) \supset R(\vv{u},\vv{v}))}
{\forall \vv{u} \exists q
R_1(\vv{u},q) \wedge R_1^L([\vv{x},p]\uparrow_\vv{y}^p P(\vv{x},\vv{y}))}
{\mbox{Procedure rule}}

\RRule{R^L(P(\vv{x},p);\downarrow_\vv{y}^p), P \mbox{ non-empty}}
{R^L(\downarrow_\vv{y}^p P(\vv{x},p))}
{\mbox{Application rule 1}}

\RRule{q(\vv{v}) \supset R(\vv{v},q)}
{R^L([\vv{y},p]\downarrow_\vv{y}^p)}
{\mbox{Application rule 2}}

\RRule{
\begin{array}{c}
{\forall \vv{x}'u'\Sigma \vv{y}' v'R(\vv{x}',u',\vv{y}',v'),
R^L(P(\vv{x},u,\vv{y},v)), \vv{x}, u \in P^{in}, \vv{y}, v \in P^{out}} \\
{\forall \vv{x}'u' \vv{y}' v'(R(\vv{x}',u',\vv{y}',v') \supset
\exists \vv{y}'' v'' \{\vv{x}',u',\vv{y}'',v'' : P(\vv{x},u,\vv{y},v)\})} \\
{\forall \vv{x}' \vv{y}' w'(R_1(\vv{x}',\vv{y}',w') \equiv
(R(\vv{x}',w',\vv{y}',w')\wedge
\forall z' \forall \vv{y}''(R(\vv{x}',z',\vv{y}'',z') \supset z' \geq_L w')))}
\end{array}}
{
R_1^L(\mu(u,v,P(\vv{x},u,\vv{y},v),\geq_L))}
{\mbox{Least fixpoint rule}}

\noindent
The second line of the hypothesis states that $R$ does not hold on
``bad inputs'' to $P$, that is, inputs for which there is no output.
The ordering $\geq_L$ depends on $L$ and expresses the effect of
recursion in $L$.  Usually $\geq$ abbreviates $\geq_L$.

\begin{theorem}.
These inference rules are logical consequences of the axioms for
program language semantics which appear in section \ref{PL.axioms}.
\end{theorem}

\begin{proof}
The proofs for each rule follow:

\subparagraph{Underlying logic rule}
Suppose $R_1^L(P)$ and $\forall \vv{y}(R_1(\vv{y}) \supset R_2(\vv{y}))$.
By the definition of $R_1(P)$, axiom \ref{eq:progdef},
$\forall \vv{y}(\{\vv{y} : [\vv{x}]P\} \supset
 R_1(\vv{y}))$.  Since $\forall \vv{y}(R_1(\vv{y}) \supset R_2(\vv{y}))$,
$\forall \vv{y}(\{\vv{y} : [\vv{x}]P\} \supset
 R_2(\vv{y}))$.  Again by axiom \ref{eq:progdef}, $R_2^L(P)$.

\subparagraph{Universal quantification rule}
Suppose $R_k^L(P)$ for arbitrary $k$.  By the definition of
$R_k^L$, axiom \ref{eq:progdef},
$\forall \vv{y}(\{\vv{y} : [\vv{x}]P\} \supset
 R_k^L(\vv{y}))$.
Because $k$ is arbitrary,
$\forall k(\forall \vv{y}(\{\vv{y} : [\vv{x}]P\} \supset
 R_k^L(\vv{y})))$.  Because $k$ does not appear in the antecedent,
$\forall \vv{y}(\{\vv{y} : [\vv{x}]P\} \supset
 \forall k R_k^L(\vv{y}))$.
By axiom \ref{eq:progdef},
$\forall k R_k^L(P)$.

\subparagraph{Variable renaming rule}
Suppose $R^L([\vv{x}]P(\vv{x}))$ and $\Theta$ is a variable renaming.
Suppose also that $\{\vv{y} : [\vv{x}\Theta]P(\vv{x}\Theta)\}$.
By the variable renaming axiom,
$\{\vv{y} : [\vv{u}]P(\vv{u})\} \supset
\{\vv{y} : [\vv{u}\Theta^{-1}]P(\vv{u}\Theta^{-1})\}$ because $\Theta^{-1}$
is also a variable renaming.
Letting $\vv{u}$ be $\vv{x}\Theta$, 
from the assumption  $\{\vv{y} : [\vv{x}\Theta]P(\vv{x}\Theta)\}$
it follows that
$\{\vv{y} : [\vv{x}]P(\vv{x})\}$.  Since
$R^L([\vv{x}]P(\vv{x}))$, $R(\vv{y})$.
Therefore $\{\vv{y} : [\vv{x}\Theta]P(\vv{x}\Theta)\}$
implies $R(\vv{y})$.  By the definition of $R^L$,
$R^L([\vv{x}\Theta]P(\vv{x}\Theta))$.

\subparagraph{Permutation rule 1}
Suppose $R^L([\vv{x}]P(\vv{x}))$ and $\sigma$ is an integer permutation.
Suppose also that $\{\vv{z} : [\vv{\sigma}^{-1}(\vv{x})] P(\vv{x})\}$.
By the permutation axiom,
$\{\vv{\sigma}(\vv{z}) : [\vv{x}] P(\vv{x})\}$.
Since $R^L([\vv{x}]P(\vv{x}))$, $R(\vv{\sigma}(\vv{z}))$ holds, or,
$R\sigma(\vv{z})$.
Thus $\{\vv{z} : [\vv{\sigma}^{-1}(\vv{x})] P(\vv{x})\}$ implies
$R\sigma(\vv{z})$.
Therefore
$R\sigma^L([\vv{\sigma}^{-1}(\vv{x})]P(\vv{x}))$.

\subparagraph{Permutation rule 2}
Suppose $R^L([\vv{x}]P(\vv{x}))$ and $\sigma$ is an integer permutation.
By permutation rule 1, $R\sigma^L([\vv{\sigma}^{-1}(\vv{x})]P(\vv{x}))$.
By the special case of the variable renaming axiom,
$R\sigma^L([\vv{x}]P(\vv{\sigma}(\vv{x})))$.

\subparagraph{Equality rule}
Suppose $R^L(P(\vv{x}))$, $x_i \equiv x_j$, and $R_1(\vv{y}) \equiv (R(\vv{y}) \wedge
y_i=y_j)$.  Recall that $P(\vv{x})$ abbreviates
$[\vv{x}]P(\vv{x})$.
Suppose also that $\{\vv{y} : [\vv{x}]P\}$.
Since $R^L(P(\vv{x}))$, $R(\vv{y})$ holds.
By the equality axiom,
$(\{\vv{y} : [\vv{x}]P\} \wedge x_i \equiv x_j) \supset y_i = y_j$.
Therefore $y_i=y_j$.  
Since $R_1(\vv{y}) \equiv (R(\vv{y}) \wedge
y_i=y_j)$, $R_1(\vv{y})$ holds.
Therefore $\{\vv{y} : [\vv{x}]P\}$ implies $R_1(\vv{y})$.
Therefore $R_1^L(P(\vv{x}))$.

\subparagraph{Substitution rule, special case}
Suppose $R^L([u,v,\vv{x}]P(u,v,\vv{x}))$ and $\sigma$ is an integer function
such that $\vv{\sigma}$
does not identify data variables of $\vv{x}$ and
such that $\sigma(i)=i$ for $i \neq 2$ and $\sigma(2)=1$.
Then $R\sigma(u',v',\vv{x}')=R(u',u',\vv{x}')$.  It is necessary
to show $R\sigma^L([u,v,\vv{x}]P\vv{\sigma}(u,v,\vv{x}))$, that is,
$R\sigma^L([u,v,\vv{x}]P(u,u,\vv{x}))$.
Suppose $\{u',v',\vv{x}' : [u,v,\vv{x}]P(u,u,\vv{x})\}$.
It is necessary to show $R\sigma(u',v',\vv{x}')$.
By the correspondence axiom,
$\{u',u',\vv{x}' : [u,u,\vv{x}]P(u,u,\vv{x})\}$ if $\sigma$ is output
injective.  By definitional independence,
$\{u',u',\vv{x}' : [u,v,\vv{x}]P(u,v,\vv{x})\}$ if $u$ and $v$ are not
distinct data variables.
Because $R^L([u,v,\vv{x}]P(u,v,\vv{x}))$, $R(u',u',\vv{x}')$.
Therefore $R\sigma(u',v',\vv{x}')$.
Hence $R\sigma^L([u,v,\vv{x}]P(u,u,\vv{x}))$.
Therefore
$R\sigma^L([u,v,\vv{x}]P\sigma(u,v,\vv{x}))$.

\subparagraph{Substitution rule, general case}
Suppose $R^L([\vv{x}]P(\vv{x}))$ and $\sigma$ is an integer
permutation
such that $\vv{\sigma}$
does not identify data variables of $\vv{x}$.
Then $R\sigma^L([\vv{x}]P\sigma(\vv{x}))$ by combining
permutation rule 1 and the variable renaming rule.  Now, let $\sigma$
be an arbitrary integer function from $\{1, \dots, n\}$ to $\{1, \dots,
n\}$ where $\vv{x}$ has $n$ components.  Then $\sigma$ can be
expressed as the composition $\sigma_1\sigma_2 \dots \sigma_k$ where
the $\sigma_i$ are permutations and functions as in the preceding
special case.  For each such $\sigma_i$, $R\sigma_1\sigma_2 \dots
\sigma_{i-1}^L([\vv{x}]P\sigma_1\sigma_2 \dots \sigma_{i-1}(\vv{x}))$
implies $R\sigma_1\sigma_2 \dots \sigma_i^L([\vv{x}]P\sigma_1\sigma_2
\dots \sigma_i(\vv{x}))$.  By combining all these implications,
$R^L([\vv{x}]P(\vv{x}))$ implies $R\sigma^L([\vv{x}]P\sigma(\vv{x}))$.

\subparagraph{Correspondence rule}
Suppose $R^L([\vv{x}])P$.
It is necessary to show $R_1^L([\vv{y}]P)$.
For this, suppose $\{\vv{y}' : [\vv{y}]P\}$.
Let $f$ and $g$ be such that
$\vv{g}(\vv{y})=\vv{y}'$ and
$\forall u \in FV(P) \cup ([\vv{x}]P)^{data}
(f(u)=g(u))$.
Then $\{\vv{g}(\vv{y}) : [\vv{y}]P\}$.
By the correspondence axiom, with $f$ and $g$ interchanged,
$\{\vv{f}(\vv{x}) : [\vv{x}]P\}$.  Because $R^L([\vv{x}]P)$,
$R(\vv{f}(\vv{x}))$.  By the definition of $R_1$,
$R_1(\vv{g}(\vv{y}))$, that is,
$R_1(\vv{y}')$.

\subparagraph{Definitional independence rule}
Suppose $R^L([\vv{x}]P(\vv{x}))$ and $\Theta$ is output injective on $P$.
Suppose also that $\{\vv{z} : [\vv{x}\Theta] P(\vv{x}\Theta)\}$.
By the definitional independence axiom,
if $\Theta$ is a variable substitution that is output-injective on $P$
and does not identify two data variables,
then
$\forall \vv{z} (\{\vv{z} : [\vv{x}\Theta] P(\vv{x}\Theta)\} \supset 
\{\vv{z} : [\vv{x}] P(\vv{x})\})$.
Therefore $\{\vv{z} : [\vv{x}] P(\vv{x})\}$.
Since $R^L([\vv{x}]P(\vv{x}))$, $R(\vv{z})$ holds.
Thus $\{\vv{z} : [\vv{x}\Theta] P(\vv{x}\Theta)\}$ implies
$R(\vv{z})$.
Therefore
$R^L([\vv{x}\Theta]P(\vv{x}\Theta))$.

\subparagraph{New variable rule}
Suppose
$R^L([\vv{x},\psi]P)$, $y$ is a new data variable for $P$, and
$R(\vv{u},v) \wedge R(\vv{u},w) \supset R_1(\vv{u},v,w)$.
Suppose also that $\{\vv{u},v,w : [\vv{x},y,\psi] P\}$.
By the new variable axiom,
$\{\vv{u},v : [\vv{x},\psi] P\}$ and
$\{\vv{u},w : [\vv{x},\psi] P\}$.  Because
$R^L([\vv{x},\psi]P)$,
$R(\vv{u},v)$ and $R(\vv{u},w)$.  By the above implication,
$R_1(\vv{u},v,w)$.  Therefore
$R_1^L([\vv{x},y,\psi]P)$.

\subparagraph{Composition rule}
Suppose $R_1^L(P_1(\vv{y}))$, $R_2^L(P_2(\vv{z}))$, 
$prec(\phi,P_1,P_2,f,g)$,
and
$\forall \vv{f} \vv{g}(R_1(\vv{f}(\vv{y})) \wedge R_2(\vv{g}(\vv{z})) \supset
R(\phi_{P_1,P_2}(\vv{f},\vv{g})(\vv{y} \circ \vv{z})))$.
Suppose also that $\{\vv{y}'\circ \vv{z}' :
P_1(\vv{y}); P_2(\vv{z})\}$.  Then there must be a
semantic function $h$ such that
$\{\vv{h}(\vv{y} \circ \vv{z}) :
P_1(\vv{y}); P_2(\vv{z})\}$.
By the sequential composition axiom
\ref{eq:composition}, there are semantic functions $f$ and $g$ such that
$\{\vv{f}(\vv{y}) : P_1(\vv{y})\}$ and $\{\vv{g}(\vv{z}) :
P_2(\vv{z})\}$ and $h = \phi_{P_1,P_2}(f,g)$
and $prec(\phi,P_1,P_2,f,g)$.
From $R_1^L(P_1(\vv{y}))$ and $R_2^L(P_2(\vv{z}))$ it
follows that $R_1(\vv{f}(\vv{y}))$ and $R_2(\vv{g}(\vv{z}))$.
From
$\forall \vv{f} \vv{g}(R_1(\vv{f}(\vv{y})) \wedge R_2(\vv{g}(\vv{z})) \supset
R(\phi_{P_1,P_2}(\vv{f},\vv{g})(\vv{y} \circ \vv{z})))$
it follows that
$R(\phi_{P_2,P_2}(\vv{f},\vv{g})(\vv{y} \circ \vv{z})))$
Therefore
$R(\vv{h}(\vv{y} \circ \vv{z})))$,
so $R(\vv{y}'\circ \vv{z}')$.
Therefore $\{\vv{y}' \circ \vv{z}' : P_1(\vv{y});
P_2(\vv{z})\} \supset R(\vv{y}',\vv{z}')$.  By axiom \ref{eq:progdef},
$R^L(P_1(\vv{y}); P_2(\vv{z}))$.

\subparagraph{Conditional rule}
Suppose $R_1^L(P_1(\vv{y}))$, $R_2^L(P_2(\vv{z}))$, 
$\vv{y} \subseteq FV(P_1)$, $\vv{z} \subseteq FV(P_2)$,
and 
$\forall x' \vv{y}' \vv{z}'(x' \wedge R_1(\vv{y}')) \vee 
(\neg x' \wedge R_2(\vv{z}')) \supset
R(x',\vv{y}',\vv{z}')$.
 Suppose also that $\{x',\vv{y}',\vv{z}' : x ? P_1(\vv{y}) ? P_2(\vv{z})\}$.
By the conditional axiom
\ref{eq:composition}, 
$(x'={\bf true} \wedge \{\vv{y}':[\vv{y}]P_1(\vv{y})\}) \vee
 (x'={\bf false} \wedge \{\vv{z}':[\vv{z}]P_2(\vv{z})\})$.
From $R_1^L(P_1(\vv{y}))$ and $R_2^L(P_2(\vv{z}))$ it
follows that
$(x'={\bf true} \wedge R_1(\vv{y}')) \vee
 (x'={\bf false} \wedge R_2(\vv{z}'))$.
From 
$\forall x' \vv{y}' \vv{z}'(x' \wedge R_1(\vv{y}')) \vee 
(\neg x' \wedge R_2(\vv{z}')) \supset
R(x',\vv{y}',\vv{z}')$.
it follows that
$R(x',\vv{y}',\vv{z}')$.  Therefore $\{x',\vv{y}',\vv{z}' : x?P_1(\vv{y})?
P_2(\vv{z})\} \supset R(x',\vv{y}',\vv{z}')$.  By axiom \ref{eq:progdef},
$R^L([x,\vv{y},\vv{z}]x ? P_1(\vv{y}) ? P_2(\vv{z}))$.

\subparagraph{Output deletion rule}
Suppose $R^L([\vv{u},x,\vv{v}]P)$ and
$R_1(\vv{u},\vv{v}) \equiv \exists x R(\vv{u},x,\vv{v})$.
Suppose $\{\vv{u}',\vv{v}' : [\vv{u},\vv{v}]\exists xP\}$.
By the deleting output axiom,
\begin{equation}
\forall \vv{u}'\vv{v}'(\exists x'\{\vv{u}',x',\vv{v}' :
[\vv{u},x,\vv{v}]P\} \equiv \{\vv{u}',\vv{v}' : [\vv{u},\vv{v}]\exists xP\})
\end{equation}
Therefore $\exists x'\{\vv{u}',x',\vv{v}' :
[\vv{u},x,\vv{v}]P\}$.
Because $R^L([\vv{u},x,\vv{v}]P)$, it follows that
$\exists x' R(\vv{u}',x',\vv{v}')$.
Therefore by definition of $R_1$, $R_1(\vv{u}',\vv{v}')$.  Because
$\{\vv{u}',\vv{v}' : [\vv{u},\vv{v}]\exists xP\}$
implies $R_1(\vv{u}',\vv{v}')$,
therefore
$R_1^L([\vv{u},\vv{v}]\exists xP)$.

\subparagraph{Procedure rule}
Suppose
$R^L(P(\vv{x},\vv{y}))$ and $R_1(\vv{u},q) \equiv \forall \vv{v}
(q(\vv{v}) \supset R(\vv{u},\vv{v}))$.  It is necessary to show
$\forall \vv{u} \exists q R_1(\vv{u},q)$ and
$R_1^L([\vv{x},p]\uparrow_\vv{y}^p P(\vv{x},\vv{y}))$.
By the procedure operator axioms,
\begin{equation}
\label{procedure1.copy}
\forall \vv{u} \exists q \{\vv{u},q : [\vv{x},p] \uparrow^p_{\vv{y}}
P(\vv{x},\vv{y})\}
\end{equation}
and
\begin{equation}
\label{procedure2.copy}
\forall p q \vv{u} (\{\vv{u},q : [\vv{x},p] \uparrow^p_{\vv{y}}
P(\vv{x},\vv{y})\} \supset \forall \vv{v} (\{\vv{u},\vv{v} :
[\vv{x},\vv{y}]P(\vv{x},\vv{y})\} \equiv q(\vv{v})))
\end{equation}
where $p$ is a new variable.  From equations \ref{procedure1.copy} and
\ref{procedure2.copy} it follows that
\begin{equation}
\label{procedure3.proof}
\forall \vv{u} \exists q \forall \vv{v} (\{\vv{u},\vv{v} :
[\vv{x},\vv{y}]P(\vv{x},\vv{y})\} \equiv q(\vv{v})).
\end{equation}
By the definition of $R^L(P(\vv{x},\vv{y}))$,
it follows that $\{\vv{u},\vv{v} : [\vv{x},\vv{y}]P(\vv{x},\vv{y})\} \supset
R(\vv{u},\vv{v})$.  From this and equation \ref{procedure3.proof}
it follows that $\forall \vv{u} \exists q \forall \vv{v} (q(\vv{v})
\supset R(\vv{u},\vv{v}))$, and by the definition of $R_1$ this
implies $\forall \vv{u} \exists q R_1(\vv{u},q)$.  From
equation \ref{procedure2.copy} and the fact that $R^L(P(\vv{x},\vv{y}))$
and by the definition of $R_1$ it follows that
\begin{equation}
\forall p q \vv{u} (\{\vv{u},q : [\vv{x},p] \uparrow^p_{\vv{y}}
P(\vv{x},\vv{y})\} \supset R_1(\vv{u},q).
\end{equation}
Therefore $R_1^L([\vv{x},p]\uparrow_\vv{y}^p P(\vv{x},\vv{y}))$.

\subparagraph{Application rule 2}

Suppose 
$q(\vv{v}) \supset
R(\vv{v},q)$.
It is necessary to show
$R^L([\vv{y},p]\downarrow_\vv{y}^p)$.
By the second application axiom,
$\forall \vv{v}q(\{\vv{v},q : [\vv{y},p]\downarrow^p_{\vv{y}}\}
\equiv q(\vv{v}))$.
Because 
$q(\vv{v}) \supset
R(\vv{v},q)$,
$\forall \vv{v}q(\{\vv{v},q : [\vv{y},p]\downarrow^p_{\vv{y}}\}
\supset R(\vv{v},q))$.
Therefore $R^L([\vv{y},p]\downarrow_\vv{y}^p)$.

\subparagraph{Least fixpoint rule}
Suppose
\begin{equation}
\forall \vv{x}'u'\Sigma \vv{y}' v'R(\vv{x}',u',\vv{y}',v'),
R^L(P(\vv{x},u,\vv{y},v)), \vv{x}, u \in P^{in}, \vv{y}, v \in P^{out}
\label{fixproof.1}
\end{equation}
\begin{equation}
{\forall \vv{x}'u' \vv{y}' v'(R(\vv{x}',u',\vv{y}',v') \supset
\exists \vv{y}'' v'' \{\vv{x}',u',\vv{y}'',v'' : P(\vv{x},u,\vv{y},v)\})} \\
\label{fixproof.1.5}
\end{equation}
\begin{equation}
\forall \vv{x}' \vv{y}' w'(R_1(\vv{x}',\vv{y}',w') \equiv
(R(\vv{x}',w',\vv{y}',w')\wedge
\forall z' \forall \vv{y}''(R(\vv{x}',z',\vv{y}'',z') \supset z' \geq w')))
\label{fixproof.2}
\end{equation}
It is necessary to show $R_1^L(\mu(u,v,P(\vv{x},u,\vv{y},v),\geq))$.
Assume
\begin{equation}
\{\vv{x}',\vv{y}',w' : \mu(u,v,P,\geq)(\vv{x},\vv{y},v)\}.
\label{fixproof.assumption}
\end{equation}
It is necessary to show $R_1(\vv{x}',\vv{y}',w')$, that is,
\begin{equation}
R(\vv{x}',w',\vv{y}',w')\wedge
\forall z' \forall \vv{y}''(R(\vv{x}',z',\vv{y}'',z') \supset z' \geq w').
\label{fixproof.goal}
\end{equation}
By the least fixpoint axiom, if
\begin{equation}
\forall \vv{x}' u' \Sigma \vv{y}' v' \{\vv{x}',u',\vv{y}',v' :
P(\vv{x},u,\vv{y},v)\}
\label{fixproof.3}
\end{equation}
and $\vv{y},v$ are outputs and $\vv{x},u$ are inputs of
$P$ then
\begin{eqnarray}
\{\vv{x}',\vv{y}',w' : \mu(u,v,P,\geq)(\vv{x},\vv{y},v)\} & \equiv &
\nonumber \\ (\{\vv{x}', w', \vv{y}', w' : P(\vv{x}, v, \vv{y}, v)\} &
\wedge & \forall z' \forall \vv{y}'' (\{\vv{x}', z', \vv{y}'', z' :
P(\vv{x}, u, \vv{y}, v)\} \supset z' \geq w')).
\label{fixproof.4}
\end{eqnarray}
Now, formula \ref{fixproof.3} follows from formula \ref{fixproof.1}.
For if 
$\{\vv{x}',u',\vv{y}',v' : P(\vv{x},u,\vv{y},v)\}$ and
$\{\vv{x}',u',\vv{y}'',v'' : P(\vv{x},u,\vv{y},v)\}$ then by
formula \ref{fixproof.1}, 
$R^L(P(\vv{x},u,\vv{y},v))$, so
$R(\vv{x}',u',\vv{y}',v')$ and
$R(\vv{x}',u',\vv{y}'',v'')$ hold, and from the formula
$\forall \vv{x}'u'\Sigma \vv{y}' v'R(\vv{x}',u',\vv{y}',v')$ it follows
that $\vv{y}'=\vv{y}''$ and $v' = v''$.  This proves formula
\ref{fixproof.3}.  Then
by the least fixpoint axiom,
formula \ref{fixproof.4} follows.
From formula \ref{fixproof.4} and assumption \ref{fixproof.assumption}
it follows that $\{\vv{x}', w', \vv{y}', w' : P(\vv{x}, v, \vv{y}, v)\}$,
and therefore from $R^L(P(\vv{x},u,\vv{y},v))$ it follows that
$R(\vv{x}', w', \vv{y}', w')$.  To prove formula \ref{fixproof.goal},
it is also necessary to show
\begin{equation}
\forall z' \forall \vv{y}''(R(\vv{x}',z',\vv{y}'',z') \supset z' \geq w').
\label{fixproof.5}
\end{equation}
Suppose $R(\vv{x}',z',\vv{y}'',z')$.
From formula \ref{fixproof.1.5}
it follows that formula
$\{\vv{x}', z', \vv{y}''', z'' :
P(\vv{x}, u, \vv{y}, v)\}$
holds for {\em some} $\vv{y}'''$ and $z''$, and therefore
$R(\vv{x}',z',\vv{y}''',z'')$.  Since $\vv{y}'''$ and $z''$ are unique
by formula \ref{fixproof.1}, they are equal to $\vv{y}''$ and $z'$.
Therefore $\{\vv{x}', z', \vv{y}'', z' : P(\vv{x}, u, \vv{y}, v)\}$, and
then from formula \ref{fixproof.4} it follows that $z' \geq w'$.  This
completes the proof.

\end{proof}

There is an algorithm to extract $L$ programs from proofs in
relational $PL(L)$, as follows:

\begin{Define}
The {\em $PL$ program operations} are composition $(;)$, conditional
$(?)$, variable deletion $(\exists)$, procedure $(\uparrow)$,
application $(\downarrow)$, and least fixpoint $(\mu)$.
\end{Define}

\begin{Define}
If $L$ is a $PL$-feasible language, and $P_1, P_2, \dots, P_n$ are $L$
programs, then an {\em $L$ program term} over $P_1, P_2, \dots, P_n$
is either
\begin{enumerate}
\item one of the programs $P_i$, or
\item of the form $P ; Q$, $x?P?Q$, $\uparrow^p_{\vv{x}}P$,
$\downarrow^p_{\vv{x}}P$, $\exists x P$, or $fp(x,y,P,\geq)$ where $P$
and $Q$ are $L$ program terms over $P_1, P_2, \dots, P_n$ and $x$,
$y$, $p$, and $\vv{x}$ are program variables, and where the preconditions
for these operators are satisfied.
\end{enumerate}
\end{Define}

\begin{theorem}
If $L$ is a $PL$-feasible language and there is a proof of an
assertion of the form $R^L([\vv{y}]P)$ from assertions of the form
$R_i^L([\vv{y}^i]P_i)$ in relational $PL(L)$, then $P$ is expressible
as an $L$ program term over $P_1\Theta_1$, $P_2\Theta_2$, $\dots$,
$P_n\Theta_n$ for some output injective variable substitutions
$\Theta_i$.
\end{theorem}

\begin{proof}
By induction on proof depth.
For depth 0, $P = P_i$ for some $i$, and $P_i$ is trivially an $L$ program
term over $P_1, \dots, P_n$.
Assume the theorem is true for proofs of
depth $d$.  A proof of depth $d+1$ consists of one or two proofs of
depth $d$ followed by the application of an inference rule.  By induction,
the theorem is true for the proof or proofs of depth $d$.  Then, using
the forms of the inference rules, the theorem is also true for the
proof of depth $d+1$.

It is necessary to look at each inference rule.  For the underlying
logic rule, $P$ is not altered, so the induction step holds.  For
permutation rules 1 and 2, only the variables of $P$ are renamed,
and renaming variables in an $L$ program term over
$P_1\Theta_1, \dots, P_n\Theta_n$ yields another $L$ program term over
$P_1\Theta'_1, \dots, P_n\Theta'_n$ for suitable $\Theta'_i$.  The
equality rule does not alter $P$.  The substitution rule applies an
integer function $\sigma$ to the variables of $P$.  This can
be incorporated into the $\Theta_i$ as well, but it is necessary to
check that identifying variables of the $P_i$ does not invalidate any
inference rules used to obtain $P$.  The correspondence rule does not
affect $P$, only its preceding list of variables.  The definitional
independence rule is similar to the substitution rule in its effect on
$P$.  The remaining rules (composition rule, conditional rule, output
deletion rule, procedure rule, application rule, and least fixpoint rule)
all produce $L$ program terms from $L$ program terms.

Now, the $PL$ program operations of composition, conditional, output
deletion, $\uparrow$, $\downarrow$, and $\mu$ have some preconditions,
and it is necessary to check that these preconditions still hold in
the resulting $L$ program term after applying the substitution rule
and the definitional independence rule.  The sequential composition
operator $P;Q$ requires that no variable be in $P^{out}\cap Q^{out}$
and that no data variable be in $P^{in} \cap Q^{out}$.  Assuming the
former condition is true when the $;$ operator is applied, it will
remain true because all substitutions are output injective.  The
latter condition on data variables will remain true because no
substitution identifies two data variables unless both are inputs.
The procedure operator axiom for $\uparrow^p P$ requires that $p$ be a
new variable or an input variable for $P$.  Now, $p$ is an output
variable of $\uparrow^p P$, which implies that no substitution will
identify $p$ with any other output variable (because substitutions are
output injective).  Thus any substitution $\Theta$ will only identify
$P$ with input variables, so $p$ will still be an input variable in
$P\Theta$ and the preconditions for this rule will still hold.  The
preconditions for the application axiom are similar to those for
composition, and similar reasoning applies.  The preconditions for
$\mu(u,v,P,\geq)$ state that $u$ is an input and $v$ is an output to
$P$.  Also, $u$ is not a free variable of $\mu(u,v,P,\geq)$ and $v$ is
an output.  Because of the rules for applying substitutions, $u$ and
$v$ will remain distinct, and $u$ will remain an input variable after
substitutions are applied.  $v$ will remain an output variable as
well, because the common image of an input and an output variable is
an output variable, so the preconditions for $\mu$ will continue to
hold.

\end{proof}

\begin{corollary}
\label{L.effective.corollary}
If in addition the $PL$ program operations composition, conditional,
$\exists$, $\uparrow$, $\downarrow$, and $\mu$ are effectively
computable in $L$, in the sense that an $L$ program for $P;Q$ can
effectively be obtained from $L$ programs for $P$ and $Q$, et cetera,
then an $L$ program $P$ such that $R^L([\vv{y}]P)$ is effectively computable
from the proof, given $L$ programs for $P_1$, $P_2$, $\dots$, $P_n$.
\end{corollary}

\section{Abstract Inference Rules}
The preceding inference rules permit proofs of properties of programs
in a specific programming language $L$.  It is possible to modify
these rules to obtain the system $PL^*$ that permits abstract proofs
of the existence of programs, but not in a specific language.  Such
proofs can then can be translated into programs in specific
$PL$-feasible programming languages automatically.  These abstract
rules involve assertions of the form $(\exists P) R^L(P)$ where $P$ is
a variable representing a program and $R$ is a relation on programs
and $L$ is a variable representing a $PL$-feasible language.  The form of
the proof not only guarantees that such a program $P$ exists, but also
permits a specific program to be derived from the proof, as in other
program generation systems.  It is necessary to record the list of
input and output variables for each program variable $P$ in order to
use these inference rules; rules appearing in section \ref{relational.rules}
suffice to
compute these lists for variables $P$ appearing in the conclusion of
each rule.

\RRule{(\exists P) R_1^L(P), R_1 \supset R_2}{(\exists P) R_2^L(P)}{\mbox{Abstract underlying logic rule}}

\RRule{(\exists P)R_k^L(P), k \mbox{ arbitrary }}
{(\exists P)\forall k R_k^L(P)}{\mbox{ Abstract universal quantification rule}}

\paragraph{Rules about variables}

\RRule{(\exists P) R^L([\vv{x}]P(\vv{x})), \Theta \mbox{ is a variable renaming}}
{(\exists P) R^L([\vv{x}\Theta]P(\vv{x}\Theta))}
{\mbox{Abstract variable renaming rule}}

\RRule{(\exists P) R^{L,\vv{\psi}}([\vv{x}]P(\vv{x})), \sigma \mbox{ is an integer permutation}}
{(\exists P) R\sigma^{L,\vv{\psi}}([\vv{\sigma}^{-1}(\vv{x})]P(\vv{x}))}
{\mbox{Abstract permutation rule 1}}

\RRule{(\exists P) R^L([\vv{x}]P(\vv{x})), \sigma \mbox{ is an integer permutation}}
{(\exists P) R\sigma^L([\vv{x}]P(\vv{\sigma}(\vv{x})))}
{\mbox{Abstract permutation rule 2}}

\RRule{(\exists P) R^L(P(\vv{x})), x_i \equiv x_j, R_1(\vv{y}) \equiv (R(\vv{y}) \wedge y_i=y_j)}
{(\exists P) R_1^L(P(\vv{x}))}{\mbox{Abstract equality rule}}

\RRule{
\begin{array}{c}
{(\exists P)R^L([\vv{x}]P(\vv{x})),\sigma
\mbox{ is an integer function such that }} \\
{\vv{\sigma} \mbox{ does not identify data variables of } \vv{x}}
\end{array}}
{(\exists P)R\sigma^L([\vv{x}]P(\vv{\sigma}(\vv{x}))}{\mbox{Abstract substitution rule}}

\RRule{(\exists P)R^L([\vv{x}]P), \forall f,g(
R(\vv{f}(\vv{x})) \wedge \forall u \in FV(P) \cup ([\vv{x}]P)^{data}
(f(u)=g(u))) \supset R_1(\vv{g}(\vv{y}))}
{(\exists P)R_1^L([\vv{y}]P)}
{\begin{array}{c}
{\mbox{Abstract}}\\
{\mbox{correspondence rule}}
\end{array}}

\RRule{
\begin{array}{c}
{(\exists P)R^L([\vv{x}]P(\vv{x})), \Theta \mbox{ output injective on }P}
\\
{\mbox{ and does not identify data variables of } P}
\end{array}}
{(\exists P)R^L([\vv{x}\Theta]P(\vv{x}\Theta))}{\mbox{Abstract definitional independence rule}}

\RRule{(\exists P)R^{L,\psi}([\vv{x},\psi]P), y \mbox{ a new data variable},
R(\vv{u},v) \wedge R(\vv{u},w) \supset R_1(\vv{u},v,w)}
{(\exists P)R_1^{L,\psi}([\vv{x},y,\psi]P)}{\mbox{Abstract new variable rule}}

\paragraph{Rules about program operations}

\RRule{
\begin{array}{c}
{(\exists P_1) R_1^L(P_1(\vv{y})), (\exists P_2) R_2^L(P_2(\vv{z})),}\\
{prec(\phi,P,Q,f,g),
\forall \vv{f} \vv{g}(R_1(\vv{f}(\vv{y})) \wedge R_2(\vv{g}(\vv{z})) \supset
R(\phi_{P,Q}(\vv{f},\vv{g})(\vv{y} \circ \vv{z})))}
\end{array}}
{(\exists P) R^L(P(\vv{y},\vv{z}))}{\mbox{Abstract composition
rule}}

\RRule{
\begin{array}{c}
(\exists P_1)R_1^L(P_1(\vv{y})), (\exists P_2)R_2^L(P_2(\vv{z})), 
\vv{y} \subseteq FV(P_1), \vv{z} \subseteq FV(P_2),\\
\forall x' \vv{y}' \vv{z}'(x' \wedge R_1(\vv{y}')) \vee 
(\neg x' \wedge R_2(\vv{z}')) \supset
R(x',\vv{y}',\vv{z}')
\end{array}}
{(\exists P) R^L(P(x,\vv{y},\vv{z}))}{\mbox{Abstract conditional rule}}

\RRule{(\exists P) R^L([\vv{u},x,\vv{v}]P),
R_1(\vv{u},\vv{v}) \equiv \exists x R(\vv{u},x,\vv{v})}
{(\exists P) R_1^L(P)}{\mbox{Abstract output deletion rule}}

\RRule{(\exists P)R^L(P(\vv{x},\vv{y})), R_1(\vv{x}', q) \equiv \forall \vv{y}'
(q(\vv{y}') \supset R(\vv{x}',\vv{y}'))}
{\forall \vv{x}'\exists q
R_1(\vv{x}',q) \wedge (\exists P) R_1^L(P)}
{\mbox{Abstract procedure rule}}

\RRule{q(\vv{v}) \supset R(\vv{v},q)}
{(\exists P)R^L([\vv{y},p]P)}
{\mbox{Abstract application rule}}

\RRule{
\begin{array}{c}
{\forall \vv{x}'u'\Sigma \vv{y}'v'R(\vv{x}',u',\vv{y}',v'),}\\
{(\exists P) (R^L(P(\vv{x},u,\vv{y},v)), \vv{x}, u \in P^{in}, \vv{y},
v \in P^{out},} \\
{ \wedge \forall \vv{x}'u' \vv{y}' v'(R(\vv{x}',u',\vv{y}',v') \supset
\exists \vv{y}'' v'' \{\vv{x}',u',\vv{y}'',v'' : P(\vv{x},u,\vv{y},v)\})),}\\
{\forall \vv{x}'w'\vv{y}'(R_1(\vv{x}',\vv{y}',w') \equiv
(R(\vv{x}',w',\vv{y}',w')\wedge
\forall z(R(\vv{x}',z',\vv{y}',z') \supset z' \geq_L w')))}
\end{array}}
{(\exists P) R_1^L(P)}
{\begin{array}{c}
{\mbox{Abstract least}}\\
{\mbox{fixpoint rule}}
\end{array}}

It is possible to translate proofs in $PL^*$ into programs in
{\em any} $PL$-feasible language $M$:

\begin{theorem}
\label{abstract.proof.to.program}
There is an algorithm which, given a $PL^*$ proof of an assertion of
the form $\exists P R^L([\vv{y}]P)$ from assertions of the form
$\exists P_i R_i^L([\vv{y}^i]P_i)$ (where $L$ is a variable
representing a $PL$-feasible language), and given a $PL$-feasible
language $M$ and $M$ programs $P_i$ such that $R^M([\vv{y}^i]P_i)$,
produces an $M$-program $P$ such that $R^M([\vv{y}]P)$.
\end{theorem}

\begin{proof}
It is straightforward to translate $PL^*$ proofs into
$PL(L)$ proofs, for any $PL$-feasible language $L$, and then
apply the algorithm of corollary \ref{L.effective.corollary}.
\end{proof}

\section{Abstract programs}
Corresponding to abstract inference rules there are {\em abstract programs}
in $PL$.

\begin{Define}
An {\em abstract PL program} is either
\begin{enumerate}
\item A variable $X$, representing a program fragment, or
\item Of the form $P ; Q$, $x ? P ? Q$, $\uparrow^p_{\vv{x}}P$,
$\downarrow^p_{\vv{x}}P$, $\exists x P$, or $fp(x,y,P,\geq)$ where
$P$ and $Q$ are abstract $PL$ programs and $x$, $y$, $p$, and $\vv{x}$
are program variables.
\end{enumerate}
$L^*$ is the set of abstract $PL$ programs.  The notation $P[X_1,X_2,
\dots, X_n]$ refers to an abstract $PL$ program, where $P$ is a
composition of the $PL$ program operations $;,?,\uparrow,
\downarrow,\exists$, and $fp$ and $X_1, \dots, X_n$ is a listing of
all the variables in $P$ representing program fragments.
By contrast, $P(\vv{x})$ represents the program $P$ mentioning the
program variables $\vv{x}$.
If $P_1, P_2, \dots, P_n$ are
$L$ programs for some $PL$-feasible language $L$, then
$P[P_1, P_2, \dots, P_n]$ denotes the $L$ program term that
results from replacing all occurrences
of $X_i$ in $P[X_1, \dots, X_n]$ by $P_i$.
The program variables $\vv{x}$ can also be indicated as in
$P[P_1(\vv{x}), P_2(\vv{x}), \dots, P_n(\vv{x})]$.
\end{Define}

By theorem \ref{abstract.proof.to.program}, a $PL^*$ proof can be
converted to a $PL(L)$ proof, for any $PL$-feasible $L$, and from the
$PL(L)$ proof an $L$ program term can be obtained.  This $L$ program
term is much like an abstract $PL$ program, but it contains
substitutions on the programs $P_i$.  It is possible to eliminate
these substitutions and also the dependence on the particular proof
system $PL(L)$ or $PL^*$, as follows.

\begin{Define}
If $P(\vv{x})$ is a $PL(L)$-program having $\vv{x}$ as free variables, then
$[[P(\vv{x})]]$, the {\em semantics of} $P(\vv{x})$, is the set
$\{f : \{\overline{f}(\vv{x}) : P(\vv{x})\}\}$,
where $f$ is a function from
variables to their semantics.
\end{Define}

\begin{theorem}
\label{ops.are.denotational}
If $P$ and $Q$ are $L$ programs, then $[[P;Q]]$ is a function of
$[[P]]$ and $[[Q]]$, $[[\exists x P]]$ is a function of
$[[P]]$, $[[\uparrow^p_{\vv{x}}P]]$ is a function of
$[[P]]$, and similarly for the other $PL$ program operations.
\end{theorem}

\begin{proof}
By consideration of the definition of each operation, noting that
the semantics of the operations depend only on the semantics of the
operands.
\end{proof}

\begin{Define}
\label{define.semantic.ops}
Extend the $PL$ operations on programs to operations on their
semantics, so that $[[P ; Q]] = [[P]];[[Q]]$, $[[\exists x P]] =
\exists x [[P]]$, and so on, thus giving names to the functions in
theorem \ref{ops.are.denotational}
\end{Define}

\begin{theorem}
\label{L.program.semantics}
For every $L$ program $P[P_1, \dots, P_n]$ where $P$ is an abstract
$PL$ program
(composed of the $PL$ program operations) and $P_i$ are variables
representing $L$ programs, there is a function $f^L_P$ depending on
$P$ but not on $P_1, \dots, P_n$ such that
\begin{equation}
f^L_P([[P_1]], \dots, [[P_n]]) = [[P[P_1, \dots, P_n]]]
\end{equation}
for all $L$ programs $P_1, \dots, P_n$.
\end{theorem}

\begin{proof}
By induction on the depth of $P$, using Theorem \ref{ops.are.denotational}.
\end{proof}

\begin{theorem}
\label{abstract.program.semantics}
For every abstract
$L$ program $P[X_1, \dots, X_n]$ where $P$ is an abstract $PL$ program
(composed of the $PL$ program operations) and $X_i$ are variables
representing program fragments,
there is a function $f^*_P$ depending on
$P$ but not on $X_1, \dots, X_n$ such that
\begin{equation}
f^*_P([[X_1]], \dots, [[X_n]]) = [[P[X_1, \dots, X_n]]]
\end{equation}
for all $X_1, \dots, X_n$.
\end{theorem}

\begin{proof}
By induction on the depth of $P$, using Theorem \ref{ops.are.denotational}
and Definition \ref{define.semantic.ops}
\end{proof}

\begin{theorem}
\label{abstract.equals.L.semantics}
For any $PL$ feasible language $L$ and any $L$ program $P[P_1, \dots,
P_n]$ where $P$ is an abstract $PL$ program, the abstract program $P[X_1,
\dots, X_n]$ satisfies $f^L_P = f^*_P$.
\end{theorem}

\begin{proof}
$f^L_P([[P_1]], \dots, [[P_n]]) = [[P[P_1, \dots, P_n]]]
=f^*_P([[P_1]], \dots, [[P_n]])$ by Theorems \ref{L.program.semantics}
and \ref{abstract.program.semantics}.
\end{proof}

\begin{corollary}
Suppose that $P[X_1, \dots, X_n]$ is an abstract $PL$ program and
$A_1, \dots, A_n, A$ are assertions such that for all program
fragments $X_1, \dots, X_n$, $A_1([[X_1]]) \wedge \dots \wedge
A_n([[X_n]]) \supset A([[P[X_1, \dots, X_n]]])$.  Then for any $PL$
feasible language $L$ and any $L$ programs $P_1, \dots, P_n$ of
appropriate sorts, $A_1([[P_1]]) \wedge \dots \wedge A_n([[P_n]])
\supset A([[P(P_1, \dots, P_n)]])$.
\end{corollary}

\begin{proof}
From the hypothesis $A_1([[X_1]]) \wedge \dots \wedge A_n([[X_n]]) \supset
A([[P[X_1, \dots, X_n]]])$ and Theorem \ref{abstract.program.semantics}
it follows that
$A_1([[X_1]]) \wedge \dots \wedge A_n([[X_n]]) \supset A(f^*_P([[X_1]], \dots,
[[X_n]]))$.  From Theorem \ref{abstract.equals.L.semantics}
it follows that
$A_1([[X_1]]) \wedge \dots \wedge A_n([[X_n]]) \supset A(f^L_P([[X_1]], \dots,
[[X_n]]))$.  From Theorem 
\ref{L.program.semantics} it follows that $A_1([[P_1]]) \wedge \dots
\wedge A_n([[P_n]]) \supset A([[P[P_1, \dots, P_n]]])$.
\end{proof}

This yields the following method for constructing $L$ programs satisfying
a specification:
\begin{enumerate}
\item Construct an abstract $PL$ program $P[X_1,\dots,X_n]$.
\item Show that $P$ satisfies the specification $A_1([[X_1]]) \wedge \dots \wedge
A_n([[X_n]]) \supset A([[P[X_1, \dots, X_n]]])$.
\item Choose $L$ programs $P_1, \dots, P_n$.
\item Show that these programs satisfy $A_1([[P_1]]), \dots, A_n([[P_n]])$.
\item Conclude that the $L$ program $P[P_1, \dots, P_n]$ satisfies the
specification $A([[P[P_1, \dots, P_n]]])$.
\end{enumerate}

This shows that one can construct abstract $PL$ programs satisfying a
specification, and from them one can construct $L$ programs satisfying
the specification, for any $PL$-feasible language $L$.  $L^*$
programs are somewhat similar to ``pseudocode'' descriptions of
algorithms found in textbooks, but unlike pseudocode, $L^*$ programs
have a formal syntax and semantics, which permit programs to be
verified.  It would of course be possible to verify a program $P$ in
some particular language such as C and translate $P$ to other
languages $L$.  Why is $L^*$ any better for this purpose?  The syntax
and semantics of $L^*$ are simple, making it easier to write such a
translator and the translator is more likely to produce efficient code
in $L$.  It is also easier to verify $L^*$ programs than C programs.

Another possibility would be to verify a program in lambda calculs or
$\mu$ calculus or some other language with a simple syntax and
semantics and translate this program into other languages.  An
advantage of $L^*$ is that it has features to guide the translation,
such as the distinction between procedures and data, the use of $;$ to
signify sequential composition, the use of $\exists$ to signify
variable declarations, the use of conditionals, and so on.  This means
that in the abstract program one can give guidance about how the
algorithm should be expressed to gain efficiency.

In fact, an abstract program can be considered as a way to formally
describe algorithms.  A description of an algorithm in a particular
programming language gives extraneous details related to the
programming language syntax but not to the algorithm.  A pseudocode
description of algorithms as found in textbooks does not have a
precise syntax and semantics.  Turing machine descriptions also
contain extraneous details and lack abstraction and do not capture the
efficiency of data structures.  Pure functional languages without
destructive assignments do not permit an imperative programming style,
which can lead to inefficiency.  More abstract notations such as
lambda calculus and $\mu$ calculus give too little guidance concerning
efficient code generation, which generally requires destructive
modification of data structures, side effects, and conditional
statements, and are difficult to translate efficiently into more
conventional programming languages.  Thus $PL$ is abstract enough to
avoid extraneous details about syntax but not too abstract to express
program features that have a major influence on efficiency.

The emphasis of $PL$ is not so much the automatic construction of
programs or even automatic proofs of their correctness, but rather the
ability to write abstract proofs or programs that can be translated
into a wide variety of other languages, to avoid the necessity of
writing the same program over and over again in different languages.
Probably it would be most efficient for the abstract programs to be
coded by humans and stored in a library.  It does not appear feasible
to construct complex programs by automatic program generation methods
in most cases.  The $PL$ approach permits a reduction of programmer
effort even in the absence of automatic program generation.
The system $PL$ can even be
used without formal proofs of correctness; the programs $P_i$ can be
verified to satisfy the assertions $A_i$, or this can just be checked
by testing, to gain some measure of reliability without a formal
proof.  In fact, it is not even necessary to know that the language
$L$ is $PL$-feasible;  this can be verified in a large number of cases
by testing, to gain some confidence in the reliability of the programs.

Abstract programs may be parameterized.  For example, the precision of
floating point operations may be a parameter.  If this precision is too
high, then the abstract program may not translate into as many languages.
Another example of a parameter might be the length of character
strings.  If different languages implement different length character
strings, then depending on the values of this parameter, the abstract
program would translate into a different set of languages.  However,
if the abstract program is correct regardless of the parameter values,
then the $L$ programs resulting from it will also be correct for all
values of the parameters.

The abstract programs are not necessarily easy to read or understand,
although readability is easier for the textual syntax.
Here is an abstract program for factorial:

\[\mu(f,g,\uparrow^g_{n,v}(\exists t w x y z
\downarrow^0_t;\downarrow^=_{ntw};w?\downarrow^1_v?\downarrow^1_x;
\downarrow^-_{nxy};\downarrow^f_{yz};\downarrow^*_{nzv}),\geq)\]

\noindent
Another approach is to allow the defined symbol $g$ to be one ($f$)
that already appears, yielding the following simpler program:

\[\uparrow^f_{n,v}(\exists t w x y z
\downarrow^0_t;\downarrow^=_{ntw};w?\downarrow^1_v?\downarrow^1_x;
\downarrow^-_{nxy};\downarrow^f_{yz};\downarrow^*_{nzv})\]

\noindent
This program has the following $PL$ textual syntax:

\begin{tabular}{l}
\ \ \\
{\bf proc} $f(n,v)$;\\
\ \ {\bf var} $t,w,x,y,z$;\\
\ \ {\bf call} $0(t);$\\
\ \ {\bf call} $=(n,t,w);$\\
\ \ {\bf if} $w$ {\bf then} {\bf call} $1(v)$ {\bf else} {\bf call} $1(x)$; {\bf call} $-(n,x,y)$; {\bf call} $f(y,z)$; {\bf call} $*(n,z,v)$ {\bf fi}\\
{\bf end} $f$;
\end{tabular} \\

Here $f(n,v)$ is a procedure with input $n$ and output $v$,
$0(t)$ sets $t$ to zero, $=(n,t,w)$ sets $w$ to {\bf true} if $n=t$,
{\bf false} otherwise, $1(v)$ sets $v$ to one, $-(n,x,y)$ sets
$y$ to $n-x$, and $*(n,z,v)$ sets $v$ to $n*z$.
If $w$ is true
then $v$ is set to $1$ else $x$ is set to $1$, $y$ is set to $n-x$,
$f(y,z)$ is called, and $v$ is set to $n*z$.  The $\mu$ operator
is not necessary in this case.  In fact, for many $PL$ feasible $L$,
it is never necessary to use $\mu$, because
$\mu(u,v,P[u,v,\vv{x}],\geq) = P[v,v,\vv{x}]$.

The abstract programs could be made more abstract in various ways,
such as making them polymorphic.

As an example of the use of data, consider the following program to
update all elements of an array of length $n$:

\begin{tabular}{l}
\ \ \\
1 {\bf proc} {\bf Update}($A$,$n$)\\
2 \ \ {\bf call} {\bf Update1}($A$,$n$);\\
3 \ \ {\bf if} $n$ $>$ 1 {\bf then} {\bf call} {\bf Update}($A$,$n-1$) {\bf fi}; \\
4 {\bf end} {\bf Update};
\end{tabular} \\

Here {\bf Update1}($A$,$n$) returns array $A^{fin}$ with the $n^{th}$
element updated.  The variables $A$ and $n$ are data variables and {\bf
Update} and {\bf Update1} are procedures.
Without the use of data variables, one would have to recopy the whole
array to update each element.  The fact that the parameter $A$ to
{\bf Update1} is both an input and an output of the procedure avoids
this inefficiency.

In $A^*$ notation, without syntactic sugar, the above program would
be

\[\uparrow^{Update}_{An}
(\downarrow^{Update1}_{An};
\exists xyz (\downarrow^1_x;\downarrow^-_{nxy};\downarrow^=_{y0z};
z?\downarrow^{Update}_{Ay}?))
\]

One can obtain the effect of global variables as data variables that are
inputs to several procedures, as follows:

\[\uparrow^p_{\vv{x}}P;\uparrow^q_{\vv{y}}Q\]
where $u \in P^{in}\cap Q^{in}$ and $u$ is a data variable.
Input and output are essentially global data variables representing
files, and read and write statements modify these variables.

One can obtain iterative loops by recursion, or as special procedures that
are known to the compiler and that permit compilation by iteration
instead of recursion.  They can also be added as operators to $PL$,
much as the conditional operator was added.

\section{An example: Quicksort}

This section presents an example program, quicksort, as an abstract
program with an associated proof of correctness.  A sketch of a
translation of the abstract program into a quicksort program in C
follows.  This example illustrates the proof rules as well as the
importance of destructive modification of data for program efficiency.
It is not necessary to consider side effects because no operations in
the quicksort program have side effects.

The textual syntax for the Quicksort program, with some simplifications,
is the following:

\begin{tabular}{l}
\ \ \\
1 {\bf proc} {\bf Quicksort}($A$,$p$,$r$)\\
2 \ \ {\bf var} $q$;\\
3 \ \ {\bf if} $p < r$ \\
4 \ \ {\bf then} {\bf Partition}$(A,p,r,q)$;\\
5 \ \ \ \ {\bf Quicksort}$(A,p,q)$;\\
6 \ \ \ \ {\bf Quicksort}$(A,q+1,r)$\\
7 \ \ {\bf fi}\\
8 {\bf end} {\bf Quicksort}\\
\end{tabular} \\

\noindent
This corresponds to the abstract program

\[\uparrow^{Q}_{Apr}\exists q x y
(\downarrow^{lt}_{prx}; x?(\downarrow^P_{Aprq}; \downarrow^Q_{Apq};
\downarrow^{inc}_{qy}; \downarrow^Q_{Ayr}))\]

\noindent
where $Q$ is {\bf Quicksort}, $P$ is {\bf Partition},
$lt(p,r,x)$ sets $x$ to {\bf true} if $p<r$, {\bf false} otherwise,
and $inc(q,y)$ sets $y$ to $q+1$.  Define abstract programs
$P_1, P_2, P_3, P_4$, $P_5$, and $P_6$, respectively, as follows:

\[P_1(Q,A,p,q) = \downarrow^Q_{Apq}\]
\[P_2(P,A,p,q,r) = \downarrow^P_{Aprq}\]
\[P_3(P,Q,A,p,q,r) = P_2(P,A,p,q,r);P_1(Q,A,p,q)\]
\[P_4(P,Q,A,p,q,r,y) = P_3(P,Q,A,p,q,r);
\downarrow^{inc}_{qy}; \downarrow^Q_{Ayr}\]
\[P_5(P,Q,A,p,q,r,x,y) = \downarrow^{lt}_{prx}; x? P_4(P,Q,A,p,q,r,y)\]
\[P_6(P,Q) = \uparrow^{Q}_{Apr}\exists q x y P_5(P,Q,A,p,q,r,x,y)\]

If $L$ is the language C, then $P_1(Q,A,p,q)$ might correspond
to the statement ``$Q(A,p,q);$'', $P_2(P,A,p,q,r)$ might correspond to
the statement ``$P(A,p,q,r);$'', $P_3(P,Q,A,p,q,r)$ might correspond
to the sequence ``$P(A,p,q,r); Q(A,p,q);$'' of two procedure
calls, et cetera.  The program fragment $\exists q x y$ would
correspond to the declarations ``int $q,x,y;$'' in this case, assuming
$q$, $x$, and $y$ have integer sorts, but could correspond to the
statement ``float $q,x,y;$'' if $q$, $x$, and $y$ had real number
(floating point) sorts.  The C program for $P_6(P,Q)$ would be
something like ``void $Q(A,p,r)$ int $A[],p,r$; int $q,x,y$; \{
$P_5^L(P,Q,A,p,q,r,x,y)$ \}'' where $L$ is C.  The C program for
$P_6(P,R)$ would be ``void $R(A,p,r)$ int $A[],p,r$; int $q,x,y$; \{
$P_5^L(P,R,A,p,q,r,x,y)$ \}'', showing how program variables (names of
procedures or data variables) in program text can be replaced.
Quicksort programs in other languages besides C could be generated in
a similar manner.

To give a proof of correctness, define $perm(A,B)$ for two arrays
$A$ and $B$ to specify that the elements of $B$ are a permutation of
the elements of $A$, and define relations as follows:

\[R_{perm}(A) \equiv perm(A^{init},A^{fin})\]
\[R_{bdry}(x,y)\equiv
(x^{init}=x^{fin} \wedge
y^{init}=y^{fin})\]
\[R_{split}(A,p,q,r)\equiv
\forall i j ((p^{init}\leq i \leq q^{fin} \wedge
 q^{fin} < j \leq r^{init}) \supset
 A^{fin}[i] \leq A^{fin}[j])\]
\[R_{part}(A,p,q,r)\equiv R_{perm}(A) \wedge
R_{bdry}(p,r) \wedge
(p^{init}\leq q^{fin}) \wedge (q^{fin} < r^{init}) \wedge R_{split}(A,p,q,r)\]
\[R_{sort}(A,x,y)\equiv
\forall i j (x \leq i < j \leq y \supset A^{fin}[i]\leq A^{fin}[j])\]

For convenience, it helps to identify program variables with their
semantics when defining relations.  $R^L(P(\vv{x}))$ is defined to mean
$\forall \vv{y}(\{\vv{y} : P(\vv{x})\} \supset R(\vv{y}))$.  This
can also be written as
$f \in [[P(\vv{x})]] \supset R(f(x_1) \dots f(x_n))$.
Defining $R_f(x_1, \dots, x_n)$ as
$R(f(x_1), \dots, f(x_n))$, one has
$f \in [[P(\vv{x})]] \supset R_f(\vv{x})$.  It is more convenient
to give the relations $R_f$ than $R$ because $R_f$ mentions the
program variables $\vv{x}$ rather than their semantics $\vv{y}$.
It is these relations $R_f$ rather than $R$ that follow.  For
example, in order to express that
$\forall y_1 y_2(\{y_1,y_2 : P(x_1,x_2)\} \supset y_2=y_1+1)$,
one would ordinarily define $R(y_1,y_2) \equiv (y_2 = y_1+1)$, but
identifying program variables $x_1,x_2$ with their semantics
$y_1,y_2$ one specifies $R(x_1,x_2)\equiv (x_2=x_1+1)$, which is
more intuitive because $x_1$ and $x_2$ appear in $P$.

Recall that
data variable $x$ in a program fragment $P$ has semantics
$(\alpha,\beta)$ where $\alpha$ is the initial value of $x$ and
$\beta$ is the final value.
By convention,
$(\alpha,\beta)^{init}=\alpha$ and
$(\alpha,\beta)^{fin}=\beta$.  If one identifies variables with their
semantics, $\alpha = x^{init}$ and
$\beta = x^{fin}$.

Define relations $R_1, R_2, R_3, R_4$, and $R_5$ to be satisfied
by the programs $P_1, P_2, P_3, P_4$, and $P_5$, respectively, and
relation $R^k_{qs}$ as follows:

\[R_1(A,p,q)\equiv R_{perm}(A)\wedge R_{bdry}(p,q)\wedge R_{sort}(A,p^{init},
q^{init})\]
\[R_2(A,p,q,r)\equiv ((p<r) \supset R_{part}(A,p,q,r))\]
\[R_3(A,p,q,r)\equiv ((p<r) \supset
R_{part}(A,p,q,r)\wedge R_{sort}(A,p^{init},q^{fin}))\]
\[R_4(A,p,q,r)\equiv ((p<r)\supset R_1(A,p,r))\]
\[R_5(A,p,q,r)\equiv R_1(A,p,r)\]
\[R^k_{qs}(Q) \equiv \forall A'p'q'((q'-p')\leq k
\wedge Q(A',p',q')\supset R_1(A',p',q'))\]
\[R_{qs}(Q) \equiv \forall A'p'q'(Q(A',p',q')\supset R_1(A',p',q'))\]

$R_{qs}(Q)$ is the final specification for the quicksort program.  In
order to prove correctness, it is necessary to show that for all $k$,
$R^k_{qs}(Q)$ implies $R^{k+1}_{qs}(Q)$ and use induction.  For this
purpose, define relations $R_i^k$ specifying a bound on the sizes of
the modified subarrays, as follows:

\[R_1^k(A,p,q,Q) \equiv (R^k_{qs}(Q) \wedge (q - p) \leq k
\supset R_1(A,p,q))\]
\[R_2^k(A,p,q,r) \equiv ((r - p) \leq k+1
\supset R_2(A,p,q,r))\]
\[R_i^k(A,p,q,r,Q) \equiv (R^k_{qs}(Q) \wedge (r - p) \leq k+1
\supset R_i(A,p,q,r)), i = 3,4\]
\[R_5^k(A,p,q,r,Q) \equiv ((p \geq r) \supset A^{init}=A^{fin}) \wedge
(R^k_{qs}(Q) \wedge (r - p) \leq k+1
\supset R_5(A,p,q,r))\]

The formula $R_1^{k,L}([A,p,q,Q]P_1)$ follows directly from the
application axiom, because $Q(A,p,q)$ holds, essentially, and because
$R^k_{qs}(Q)$ holds.  One then shows that if $R_2^L([A,p,q,r]P_2)$,
then $R_3^{k,L}$ $([A,p,$ $q,r,Q]P_3)$, $R_4^{k,L}([A,p,q,r,Q]P_4)$, and
$R_5^{k,L}([A,p,q,r,Q]P_5)$.  (For convenience the variables $x$,
$y$, and $P$ ({\bf Partition}) are omitted.)  
Because $k$ is arbitrary, one has
$(\forall k R_5^k)^L([A,p,q,r,Q]P_5)$ by the universal quantification rule,
and $\forall k R_5^k(A,p,q,r,Q) \equiv$
$((p \geq r) \supset A^{init}=A^{fin}) \wedge
\forall k(R^k_{qs}(Q) \wedge (r - p) \leq k+1
\supset R_5(A,p,q,r))$.
From this, using the procedure rule, one derives
$R^0_{qs}(Q) \wedge \forall k(R^k_{qs}(Q) \supset R^{k+1}_{qs}(Q))$ for the
program $[Q]P_6$.  Using the underlying logic rule and mathematical
induction, $R_{qs}(Q)$ follows.

The final program does not include code for {\bf Partition}. Any
verified program for {\bf Partition} can be inserted and will give a
correct quicksort program.  Thus the final verified code has some
flexibility.

\bibliography{paper}

\end{document}